\documentclass[journal, final]{IEEEtran}
\usepackage{graphicx}
\usepackage{steinmetz}
\usepackage{cite}
\usepackage{color}
\usepackage{mathtools}
\usepackage{multicol}
\usepackage{xcolor}
\usepackage{epstopdf}
\usepackage{amssymb}
\usepackage{amsmath}
\usepackage{hyperref}
\usepackage{multirow}
\usepackage{subcaption}
\usepackage{amsthm}
\usepackage{epstopdf}
\usepackage[font=footnotesize]{caption}

\begin{document}

\title{A Modular Small-Signal Analysis Framework for Inverter Penetrated Power
Grids: Measurement, Assembling, Aggregation, and Stability Assessment}

\author{Lingling Fan,~\IEEEmembership{Senior Member,~IEEE}, Zhixin Miao,~\IEEEmembership{Senior Member, ~IEEE}

\thanks{Lingling Fan, and Zhixin Miao are with the Department of Electrical Engineering, University of South Florida, Tampa, FL 33420  USA e-mail: { \small\texttt{linglingfan@usf.edu}}.}
} \maketitle

\begin{abstract}
Unprecedented dynamic phenomena may appear in power grids due to higher and higher penetration of
inverter-based resources (IBR), e.g., wind and solar photovoltaic (PV). A major challenge in
dynamic modeling and analysis is that unlike synchronous generators, whose analytical models are
well studied and known to system planners, inverter models are proprietary information with black
box models provided to utilities. Thus, measurement based characterization of IBR is a popular
approach to find frequency-domain response of an IBR. The resulting admittances are essentially
small-signal current/voltage relationship in frequency domain. Integrating admittances for grid
dynamic modeling and analysis requires a new framework, namely modular small-signal analysis
framework. In this visionary paper, we examine the current state-of-the-art of dynamic modeling
and analysis of power grids with IBR, including inverter admittance characterization, the
procedure of component assembling and aggregation, and stability assessment.
We push forward a computing efficient modular modeling and analysis framework  via four
visions: (i) efficient and accurate admittance model characterization via model building and
time-domain responses, (ii) accurate assembling of components, (iii) efficient aggregation,
and (iv) stability assessment relying on network admittance matrices. Challenges of
admittance-based modular analysis are demonstrated using examples and techniques to tackle those
challenges are pointed out in this visionary paper.
\end{abstract}

\begin{IEEEkeywords}
Admittance; small-signal analysis; power grids; inverter-based resources
\end{IEEEkeywords}

\IEEEpeerreviewmaketitle

\section{Introduction}

\IEEEPARstart{M}{otivation:} High penetration of inverters introduced many new dynamic phenomena in
power grids. First severe wind farm subsynchronous resonances (SSR) due to wind farm interactions
with RLC circuits occurred in Texas in 2009 \cite{adams2012ercot}.  Similar phenomena were observed
in North China \cite{xie2017characteristic} in the years followed. In 2017, three more events were
observed in Texas \cite{southTexas}.
Oscillations were also observed in real-world wind farms with weak grid interconnections
\cite{ERCOT, liu2017subsynchronous}. For voltage source converter (VSC)-based HVDC with weak grid
interconnections, some studies have also identified stability issues
\cite{harnefors2007input,mitra2013offshore}. In microgrids, non-traditional stability issues, e.g.,
droop related stability issues, voltage stability due to low bandwidth of phase-locked-loop (PLL),
have been summarized in \cite{farrokhabadi2019microgrid}. All those stability issues call for
adequate modeling and analytical methods.

While electromagnetic (EMT) simulation in environment such as PSCAD, MATLAB/SimPowerSystems, is the
major study tool, EMT simulation mainly serves as a tool for experiments, instead of a tool for
efficient analysis. To facilitate small-signal analysis, analytical models are necessary.

\noindent {\bf Two approaches:} Currently, there are two approaches of analytic model building
characterized by how small-signal models  are obtained. In the first approach, small-signal
model is obtained as a whole. This is the practice of large-scale power grid dynamic model
building. First, a nonlinear model of the entire gird with state variables being constant at
steady-state is built. A small-signal model in the format of a linear time-invariant (LTI) system
is then obtained using numerical perturbation. An example is the power system toolbox (PST) package
developed by Cheung and Chow in the early 1990s \cite{chow1992toolbox}. This approach is a
white-box approach with the assumption that all information of the system is known.

\begin{figure}[ht!]
\vspace{-0.15in}
\begin{center}
\includegraphics[width=3.25in]{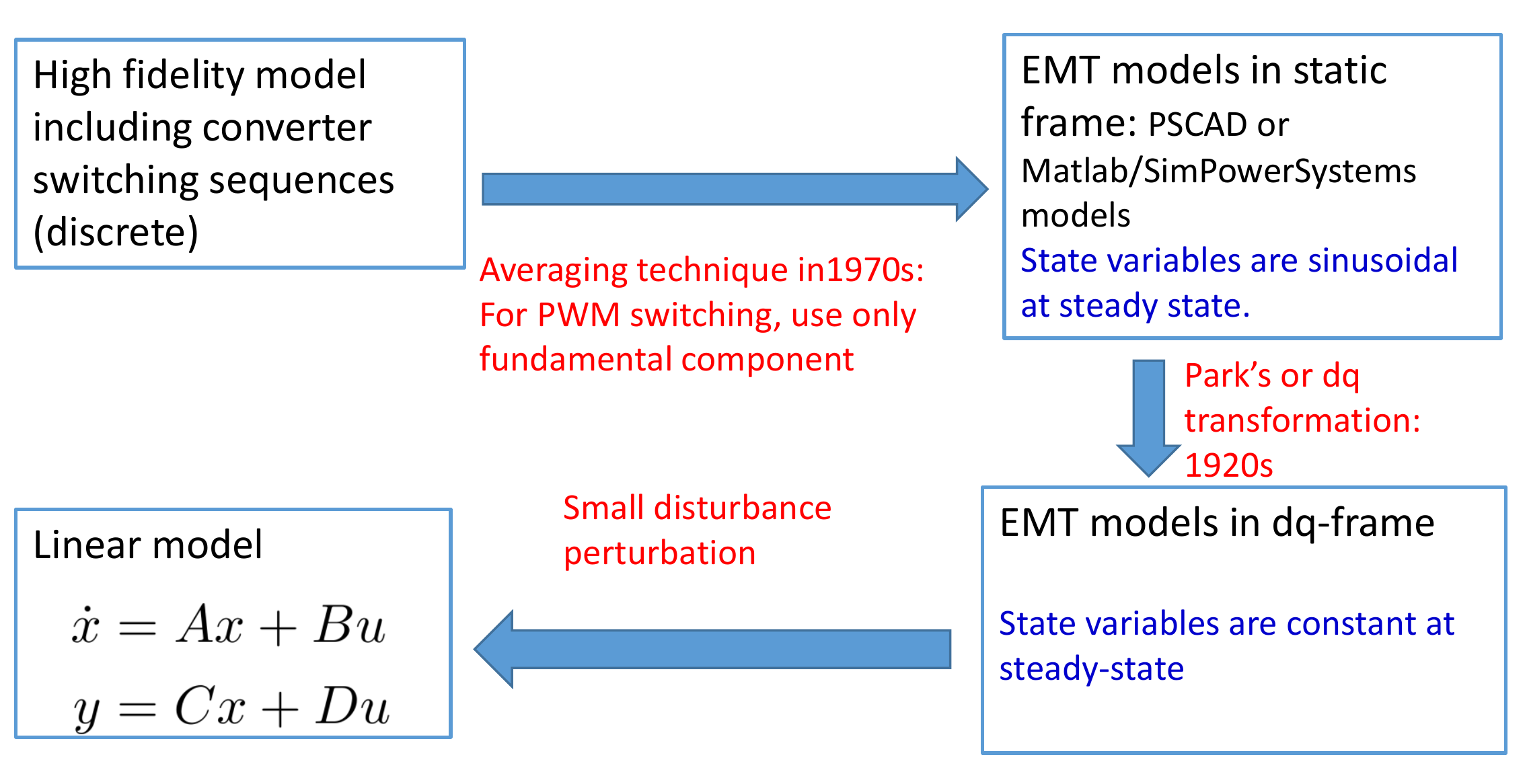}\vspace{-0.15in}
\end{center}
\caption{White-box approach: From high-fidelity models to LTI systems.}
\label{fig1:model}
\end{figure}
Indeed, to arrive at the continuous nonlinear model with state variables constant at equilibrium
points is not a trivial task. Some of the technologies are listed in Fig. \ref{fig1:model}. Imagine
a high-fidelity model of a system with converter switching details. This system has discrete state
variables. An important technique to arrive at a continuous nonlinear system is the ``averaging''
technique, whose objective is to filter out high harmonics of discrete waveforms and make signals
smooth. In the power electronics field, averaging technique was adopted in 1970s
\cite{cuk1977general}. Specifically, for discrete waveforms generated by sine pulse width
modulation (PWM), average models consider only the fundamental component of the waveforms. Many EMT
testbeds built in PSCAD and MATLAB/SimPowerSystems adopt average models to speed up computing. In
these testbeds, a voltage source converter (VSC) is usually treated as a three-phase voltage source
with controllable voltage magnitude, frequency, and phase angle. Such three-phase systems, if built
in a static reference frame, have three-phase currents and voltages as state variables. These state
variables are sinusoidal at steady-state. An important technology, $dq$-frame transformation,
converts balanced sinusoidal abc currents into dc $dq$-frame currents. In synchronous generator
modeling, the transformation is termed as Park's transformation. The 1929 paper by Park
\cite{park1929two} is ranked as one of the highest impact papers in the 20th century by the power
grid community \cite{heydt2000high}. This technique greatly speeds up dynamic
simulation. The resulting model has state variables constant at equilibrium point. 
We may apply numerical perturbation and find the linearized model. In MALTAB, the command
\texttt{linearize} or \texttt{linmod} outputs a linear model at certain initial point from a
nonlinear model. The authors' prior work on type-3 wind SSR analysis
\cite{fan2010modeling,fan2011modal}, wind in weak grids \cite{fan2018explanation,fan2018modeling,
fan2018wind, li2019wind, li2018stability} fall into the first category.

In the second approach, a small-signal model is built block by block. In power grids, the modular
approach has been adopted for synchronous generator SSR analysis by  Undrill and Kostyniak back in
1970s \cite{Undrill1976}. What's more, Undrill and Kostyniak investigated assembling components
into a network admittance for large-scale systems. Aggregation was then conducted
 to have a two-impedance system suitable for applying generalized Nyquist stability
criterion.

Another  example is the impedance model building approach popularly adopted by the power
electronics community. In the area of power electronic converters, frequency-domain impedance
model-based analysis was first used to analyze dc systems \cite{middlebrook1976input}. Sun extended
the impedance-based analysis to three-phase ac system analysis \cite{sun2009small}. Impedance
models are essentially small-signal models describing voltage and current relationship. A feature
of impedance model is that such frequency-domain response can be obtained from experiments,
 through frequency scan relying on harmonic
injection. The philosophy of harmonic injection method \cite{huang2009small, badrzadeh2012general}
is similar as frequency scan that has been used in power grid for synchronous generator SSR
assessment \cite{agrawal1979use,johansson2010comparison}.

In admittance or impedance characterization, a device is first connected to a 60 Hz
voltage source to operate at certain operating condition. To have voltage perturbation, an
additional voltage source with small magnitude at a frequency is connected in series with the
original voltage source. The current component at that frequency will be measured using Fast
Fourier Transformation (FFT) and the voltage/current phasor relationship, or impedance at this
frequency will be obtained. 

Thus, with an IBR device provided or its black-box model provided, frequency-domain impedance
measurement can be found through experiments, and further used for stability analysis. 
The concept of impedance has been introduced by the authors in \cite{fan2012nyquist,
miao2012impedance, miao2013impact} to describe a type-3 wind farm as an impedance. Impedance-based
frequency domain analysis successfully explains the effect of machine speed and network
compensation level on SSR oscillations.

\begin{figure}[ht!]
\centering
\includegraphics[width=3.2in]{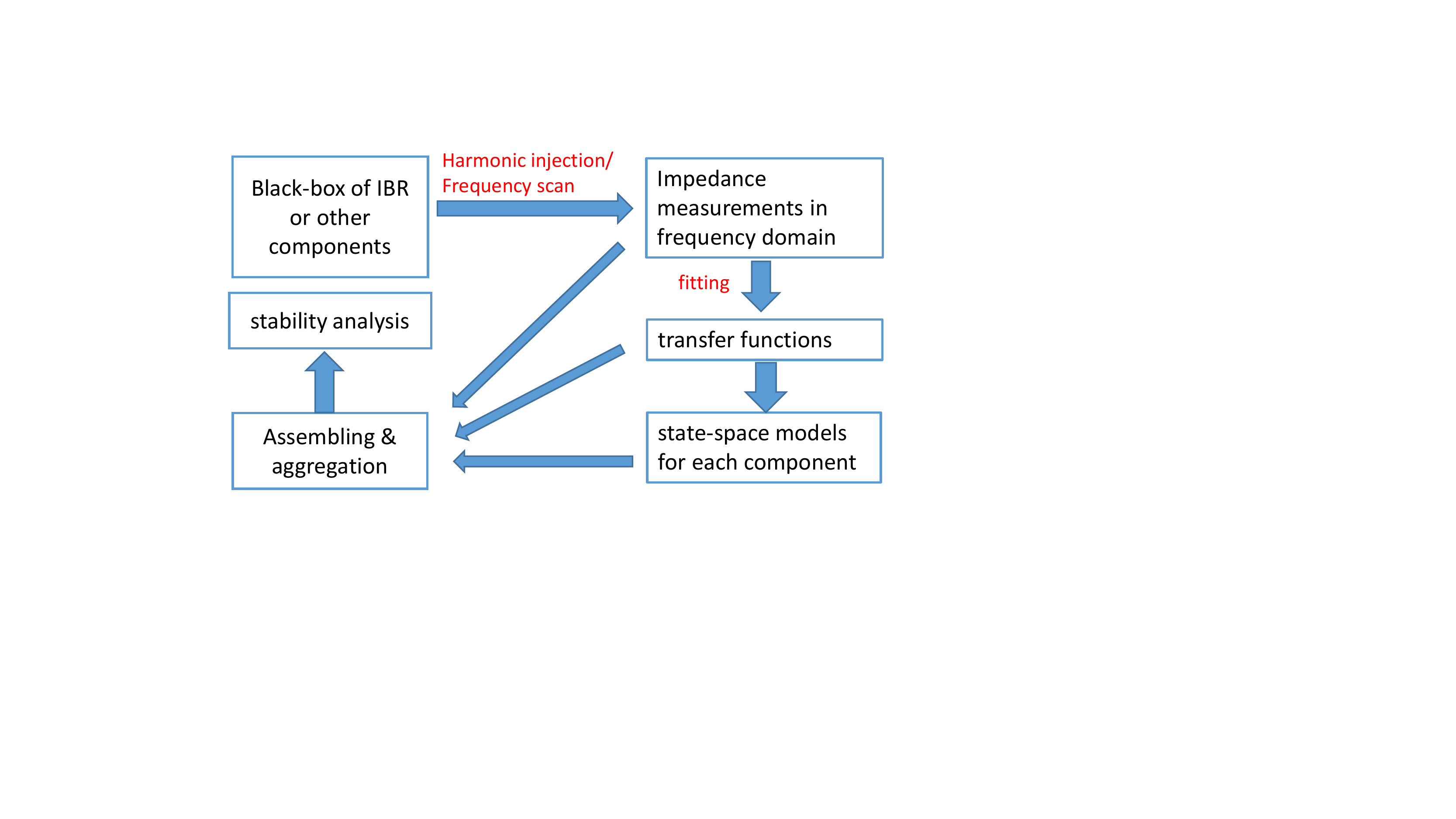}
\caption{  Black-box approach: From black boxes to LTI systems.}
\label{fig2:model}
\end{figure}

Fig. \ref{fig2:model} presents the procedure of finding a state-space model from a black-box IBR
model and further carry out stability analysis. First, admittance or impedance measurement of a
component, e.g., IBR, is obtained using harmonic injection based frequency scan method. With
measurements obtained, transfer functions may be found using vector/matrix fitting method
\cite{gustavsen1999rational}. With the frequency-domain input/output system known, time-domain
state-space model can also be found. It should be noted that the state-space model is not unique.
With each component's admittance found, the admittance of the entire system can be
formed. Stability assessment then follows.

It can be seen that the major difference between the two approaches is that in the second approach,
linearization is conducted for every component. Thus, this approach is termed as the ``modular''
approach.




\noindent {\bf Objective:} The objective of this paper is to provide visions for a scalable modular
modeling framework that is suitable for large-scale inverter penetrated power grids. To this end,
four visions will be elaborated and they are:
\begin{enumerate}
\item How to efficiently identify component admittance;
\item How to accurately assemble admittances;
\item How to efficiently find aggregation; 
\item How to analyze power system stability beyond traditional approaches. 
\end{enumerate}

\noindent {\bf Organization and Contributions:} The rest paper is organized as follows. Section II
first presents an
 overview of admittance-model based stability analysis approach. A stability criterion for a general admittance matrix is provided.
 The roots of the determinant of the admittance matrix and the system eigenvalues are equivalent. 

Section III devotes to Vision 1: How to efficiently find admittance models from nonlinear analytical models or measurement data. 
Using several examples, we first demonstrate admittance model derivation assuming all information
is known. In addition to a VSC, synchronous generator's admittance model is derived. This model is
validated against PST for its accuracy. We then propose an efficient measurement-based method
relying on time-domain responses for admittance characterization.
 This method requires only two experiments. Further, the resulting admittance model is in analytical forms instead of discrete measurements.
 For discrete measurements, further
 fitting is required to arrive at analytical forms.

 Section IV devotes to Vision 2: How to accurately assemble the component
impedances. Challenges of assembling are discussed using numerical examples. 
Section V devotes to Vision 3: How to efficiently find aggregated admittance. Technology from power
network reduction and aggregation is demonstrated in this section for efficient aggregation.

Section VI devotes to Vision 4: How to conduct stability analysis beyond the traditional
approaches. Using three examples (VSC in weak grid, type-3 wind SSR, and synchronous generator
torsional interaction), we demonstrate not only traditional approaches for stability assessment,
e.g., Nyquist Criterion, the Root-Locus method, but also methods relying on frequency-domain
admittance matrices. The new methods include eigenvalue analysis, sigma plots, and resonance mode
analysis (RMA), a technique previously used in harmonic analysis \cite{xu2005harmonic}. Finally,
Section VII concludes this paper.


\section{Admittance Model-based Small-Signal Analysis in A Nutshell}

\subsection{Analysis Based on Two Impedances}

We first use a simple system with two impedances to illustrate the stability criterion. A
two-impedance system for SSR analysis of a synchronous generator connected to an RLC circuit was
presented in a 1976 paper \cite{Undrill1976} by Undrill and Kostyniak. Generalized Nyquist
criterion is then applied for stability analysis. The two-impedance circuit is also used in
\cite{sun2009small} to illustrate impedance-based stability analysis for converter and grid
interactions.

\begin{figure}[htbp]
\vspace{-0.15in}
\centering
\includegraphics[width=2.5in]{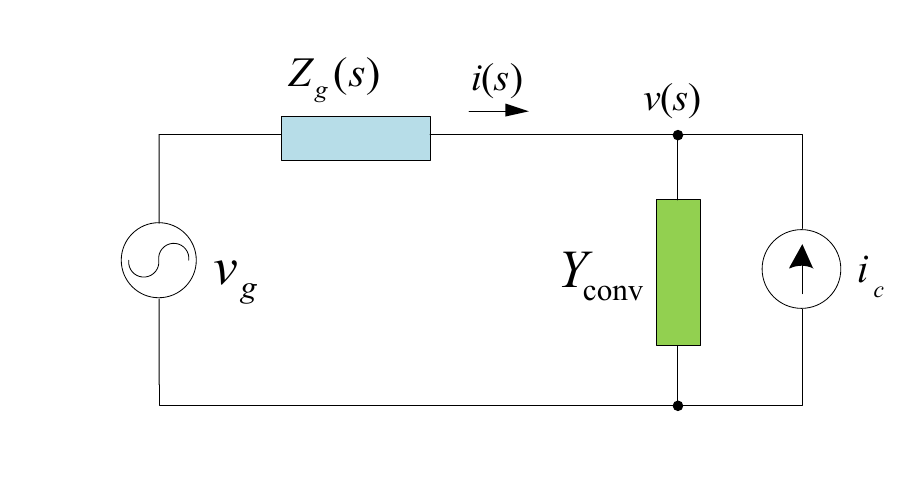}\vspace{-0.15in}
\caption{Impedance model of a converter connected to a grid.}
\label{fig:impedance}
\end{figure}

The example system in Fig. \ref{fig:impedance} represents a converter connected to a grid. The
converter is modeled in Norton representation as a current source $i_c(s)$ in parallel with an
admittance $Y_{\rm conv}(s)$. Its impedance is notated as $Z_{\rm conv}$ and $Z_{\rm conv} =
Y^{-1}_{\rm conv}$. The grid interconnection is represented as a Thevenin equivalent, a voltage
source $v_g(s)$ behind an impedance $Z_g(s)$.

The current flowing into the converter can be derived as follows.
\begin{align}
i(s) &= (Z_g+Z_{\rm conv})^{-1} (v_{g}(s)-Z_{\rm conv}i_c(s)) \notag \\
     & = \left(I +  Y_{\rm conv}Z_g\right)^{-1}(Y_{\rm conv}v_{g}(s)-i_c(s))
\end{align}
where $I$ is the identity matrix. For the above system, two assumptions are made. (i) The grid
voltage $v_{g}(s)$ is stable; (ii) the system is stable when the grid impedance $Z_g$ is zero,
i.e., $Y_{\rm conv}v_{g}(s)-i_c(s)$ is stable. The first assumption is valid for the real-world
scenarios as long as the grid voltage is within the limits. The second assumption is valid as long
as the inverter converter admittance $Y_{\rm conv}$ is stable and current order $i_c$ is stable.
For properly designed converters, the second assumption is also true.

\begin{figure}[htbp]
\centering
\includegraphics[width=2.0in]{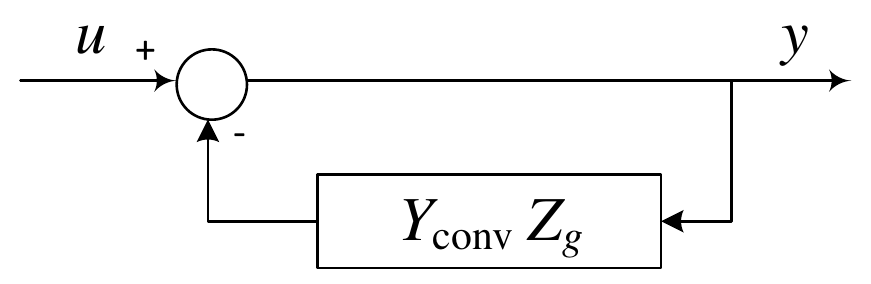}
\caption{Circuit analysis problem is converted to a feedback control problem.}
\label{fig:feedback}
\end{figure}

Therefore, for the current $i(s)$ to be stable, we only need to examine $(I + Y_{\rm
conv}Z_g)^{-1}$. This circuit analysis problem may be treated as a feedback system, as shown in
Fig. \ref{fig:feedback}, where the input $u$ is $(Y_{\rm conv}v_{g}(s)-i_c(s))$ and the output $y$
is $i(s)$.

Similarly, we may formulate a nodal equation  for analysis:
\begin{align}
v(s) = \left(Y_g +  Y_{\rm conv}\right)^{-1}(Y_g v_g(s) + i_c(s))
\label{eq:two_port}
\end{align}

%

\textbf{Stability Criterion 1: Roots of the characteristic polynomial located in the LHP.} The
system's stability is guaranteed if  the characteristic polynomial has no zero in the right half
plane (RHP) \cite{morari1989robust}, or $\det(I+ Y_{\rm conv}Z_g)=0 $ should have all roots in the
LHP.

Since $\det(I+ Y_{\rm conv}Z_g) = \det(Z_g) \det(Y_g + Y_{\rm conv})$, with the assumption that
$Z_g$ and $Y_g$ are stable, the stability criterion is equivalent to the following: the roots of
$\det(Y)$ should all be located in the LHP, where $Y$ is the total admittance and $Y = Y_g + Y_{\rm
conv}$.

Similarly, the stability criterion is equivalent to the roots of
$\det(Z_g + Z_{\rm conv})$ located at the LHP. 
This criterion is adopted in \cite{liu2018oscillatory} to compute eigenvalues of the system.


\textbf{Stability Criterion 2: Generalized Nyquist Stability Criterion or Bode Plots} More often,
open-loop analysis is used for analysis. For scalar admittances or impedances, the Root-Locus
method and Nyquist Criterion can be applied. For $2\times 2$ admittances, Generalized Nyquist
Criterion developed in 1970 \cite{MacFarlane1970} has been popularly used in the field of
frequency-domain analysis for machines and converters, e.g., \cite{belkhayat1997stability,
cespedes2014impedance}. Stability can be examined by checking the eigen loci or the Nyquist plots
of the eigenvalues of open-loop gain $Y_{\rm conv}Z_g$ or
$Z_gY_{\rm conv}$. If the eigen loci do not encircle $(-1,0)$, then the system is stable. 
In Bode plots, when phase shift happens, the gain should be less than 1 for a stable system.

%
%
%
%
%

There are other criteria used in the literature. Some are the extension of Criterion 1 and
Criterion 2 at special cases, e.g., reactance crossover adopted in \cite{cheng2013reactance,
liu2017stability}. Others, such as passivity used in \cite{harnefors2015passivity}, are suitable
for comparison, but  not accurate for stability check. Thus, these methods will not be discussed in
this paper.

\subsection{Analysis Based on a General Network Admittance}

In general, we may obtain a network admittance and establish the relationship between current
injection vector versus nodal voltage vector.
\begin{align}
V(s) &= Y(s)^{-1} I(s)
     \label{general_impedance}
\end{align}
where $Y$ is the network admittance. Capital letters are used for voltage and current to notate
vectors. Eq. \eqref{eq:two_port} is a special case of \eqref{general_impedance} where there is only
one node.

The network admittance matrix has been popularly used in power flow computing. The one used in
steady-state analysis is a special case of the network admittance presented in
\eqref{general_impedance}.

The notion of frequency-domain or $s$-domain admittance network has appeared in the literature,
e.g., \cite{semlyen1999sdomain,xu2005harmonic}. Xu \emph{et al} proposed to use the eigenvalue
information of a frequency-domain admittance network for harmonic analysis in
\cite{xu2005harmonic}. The admittance considers mainly passive elements. Semlyen indicated in
\cite{semlyen1999sdomain} that components such as generators can also be included into the
 network admittance. Semlyen also pointed out the equivalence of the system's eigenvalues and the
roots of the determinant of the admittance matrix  in \cite{semlyen1999sdomain}.

In \eqref{general_impedance}, the current injection at each node is treated as input and the nodal
voltage is treated as output. Thus, {stability of the multi-input multi-output (MIMO) system} can
be judged by the roots of the characteristic polynomial:
$\det(Y(s))=0$. {\bf The roots of $\det(Y(s))$ should be located at LHP to guarantee stability.} 
The roots and the system eigenvalues are equivalent.

Generalized Nyquist Stability criterion, however, cannot be directly applied unless the admittance
network is aggregated into a two-port system. Aggregation will be addressed in Section IV.

It is worth to notice that the stability criterion for a general network admittance has not been
popularly adopted in the literature. On reason is that the current study systems on inverters
usually have a very small size. Study on large-scale power grids with IBR penetration just starts.
Moreover, a multi-node network admittance is seen to be first reduced to a two-impedance system for
stability analysis.
This reduction approach has been adopted in \cite{Undrill1976, liu2017stability, liu2018oscillatory}. 

%

%
%



\section{Vision 1: Finding Admittance Models: White-Box Approach versus Black-Box Approach}

Two approaches are used in the literature to find admittances. The white-box approach assumes that
the information of a component is known. Admittance models may be obtained via derivation. Majority
of the current research on inverter model derivation relies on linearization at every step for
every equation, e.g., \cite{harnefors2007input, vieto2018sequence, cespedes2014impedance,
wang2017unified}. The outcome is a linear input/output model or a transfer function.

The black-box approach relies on measurement-based characterization. Frequency scan or harmonic
injection method is applied in the majority of the research, e.g., \cite{huang2009small,
badrzadeh2012general,rygg2016modified,vieto2017sequence}.  First, a testbed for measurement is
built. A small perturbation voltage is injected. FFT is then used to compute the complex Fourier
coefficients at the injected frequency. For IBRs, when measurement is conducted in the static
frame, frequency coupling phenomena may exist \cite{ren2015refined,shah2017impedance}. The phasor
of the
coupled frequency component may also need to be found and included into admittance modeling. 
On the other hand, frequency coupling phenomena have not been reported for $dq$-frame measurement
due to the fact that $dq$-frame admittances for inverters are transfer functions expressed as
rational fraction polynomials of Laplace operator $s$.

 The step of setting up the testbed to have exact operating condition is crucial for dynamics that are influenced by power system operating conditions.
Operating condition of inverters in a grid is often not taken into account in impedance modeling.
For example, in \cite{vieto2017sequence}, a type-3 wind turbine's impedance model is derived and
measured without considering power control nor operating condition. Similarly, in
\cite{wang2017unified}, a grid-connected VSC's impedance models are investigated without
considering VSC outer-loop control nor operating condition. 
The omission of operating condition and outer-loop controls renders the admittance or impedance
model not suitable for grid-level low-frequency dynamics. This point will be further emphasized in
Vision 2 Assembling.


A few research has moved forward to conduct vector/matrix fitting to obtain analytical form from
the frequency-domain measurements, e.g., \cite{rygg2016modified}. Vector/matrix fitting method was
developed  in 1999  \cite{gustavsen1999rational} and the codes are posted in the public domain.
\cite{rygg2016modified} adopts vector fitting to come up with analytical forms. Eigenvalues of the
system can be computed based on the analytical forms.

For both approaches, more \emph{efficient} approaches are desired. In power grids, nonlinear
analytical model building is a mature technology. Numerical perturbation is popularly adopted to
conduct linearization and find linear state-space models at various operating conditions. This
approach will be elaborated in Section \ref{section_whitebox}. For measurement-based
characterization, harmonic injection requires hundreds of experiments to obtain a Bode plot. On the
other hand, past research conducted by Sanchez and Chow \cite{sanchez1997computation} shows that
impulse time-domain responses can be used to find a linear state-space model or a single-input
single-output (SISO) transfer function. In Section \ref{section_blackbox}, we provide an example to
demonstrate how to use step responses to find a $2\times 2$ $dq$-frame admittance matrix. Only two
experiments are needed for this characterization. While the previous research in
\cite{sanchez1997computation} deals with measurements created from one experiment, a challenge to
tackle is to have the identification algorithm handle data created from two experiments.  A
modified algorithm that can handle data from multiple events is presented.

\subsection{White-Box Approaches}
\label{section_whitebox} Two types of admittance derivation approaches are demonstrated in this
subsection. The first approach relies on linearization conducted at every step. The second approach
is a more efficient approach where linearization is conducted on a nonlinear model.

\subsubsection{Derivation-Based Approach: Classical synchronous generator admittance
derivation}

 In this example, using the second-order
classic synchronous generator model, we show how to derive the admittance model. This is
essentially to figure out the current response of a generator with its terminal voltage perturbed.
Hence, the terminal voltage is treated as input and the current is treated as output.

The second-order model is given as follows for a generator.
\begin{align}
\dot{\delta}  = \omega_0 (\omega-1), \>
\dot{\omega}= \frac{1}{2H}(P_m - P_e -D_1(\omega-1))
\end{align}
where $\delta$ is the rotor angle in radian (defined as rotor's $q$-axis position against the
reference frame), $\omega$ is the machine speed in per unit (pu), $P_m$ and $P_e$ are the
mechanical power and the electric power in pu. $H$, $D_1$ are the inertia and damping coefficient.
The electromagnetic dynamics is ignored  and the generator is assumed as a voltage source with a
constant magnitude $E$ behind a transient reactance $X_g$. The voltage source's phase angle is the
rotor angle. Notate the generator's terminal voltage phasor as $\bar{V} = V_x + jV_y$ and the
current phasor
flowing out of the generator as $\bar{I} = I_x + jI_y$. 
The real power from the generator can be expressed as:
 $   P_e=V_x I_x + V_y I_y.$
The current phasor can be expressed by the internal voltage phasor $E\phase{\delta}$ and the
terminal voltages phasor:
    \[\bar{I}=\frac{{E}\angle{\delta}-\bar{V}}{jX_g}.\]
Hence,

\begin{align}
i_x& = \frac{E}{X_g}\sin\delta-\frac{V_y}{X_g}, \>\>
i_y = \frac{-E}{X_g}\cos\delta + \frac{V_x}{X_g}.
\label{eq:ixiy}
\end{align}

Substituting $i_x$ and $i_y$ in the power expression  leads to the following the real power
expression.
\begin{align}
    P_e=\frac{EV_x}{X_g}\sin\delta-\frac{EV_y}{X_g}\cos\delta.
\end{align}

The small perturbation expression of $\Delta P_e$ is then found:
\begin{align}
    \Delta P_e = T_x \Delta V_x + T_y\Delta V_y + T_\delta\Delta \delta
\end{align}
where
\begin{align*}
    T_x = \frac{E\sin\delta}{X_g},\>\>
    T_y = -\frac{E\cos\delta}{X_g},
    T_\delta = \frac{E(V_x\cos\delta+V_y \sin\delta)}{X_g}.
\end{align*}

It is to be noted that the expression $(V_x\cos\delta+V_y \sin\delta)$ is equivalent to
$\text{Re}(V_x + jV_y)e^{-j\delta})$, or the projection of the terminal voltage vector on the
$q$-axis of the machine that aligns with the internal voltage $E\phase{\delta}$. Thus $T_\delta$
can also be written as \[T_\delta = \frac{EV\cos(\delta-\theta_v)}{X_g},\] where $\theta_v$ is the
terminal voltage's phase angle. Hence, $T_\delta$ will be intact regardless of the reference frame
as long as the operation condition is the same.

The linear state-space model can now be found as follows.

\begin{subequations}
\begin{align}
    \begin{bmatrix}
    \Delta \dot{\delta}\\
    \Delta \dot{\omega}\\
    \end{bmatrix}
    &=
    \underbrace{\begin{bmatrix}
    0 & \omega_0 \\
    \frac{-T_\delta}{2H} & \frac{-D_1}{2H}
    \end{bmatrix}}_{A}
    \begin{bmatrix}
    \Delta \delta\\
    \Delta \omega\\
    \end{bmatrix}
    +
    \underbrace{\begin{bmatrix}
    0 & 0\\
    \frac{-T_x}{2H} & \frac{-T_y}{2H}\\
    \end{bmatrix}}_{B}
       \begin{bmatrix}
    \Delta V_x\\
    \Delta V_y\\
    \end{bmatrix}\\
\begin{bmatrix}
    \Delta {I_x}\\
    \Delta {I_y}\\
    \end{bmatrix}
    &=
    \underbrace{
    \begin{bmatrix}
    -T_y & 0 \\
     T_x & 0
    \end{bmatrix}}_{C}
    \begin{bmatrix}
    \Delta \delta\\
    \Delta \omega\\
    \end{bmatrix}
    +
    \underbrace{
    \begin{bmatrix}
    0 & \frac{-1}{X_g}\\
    \frac{1}{X_g} & 0\\
    \end{bmatrix}}_{D}
    \begin{bmatrix}
    \Delta V_x\\
    \Delta V_y\\
    \end{bmatrix}
\end{align}
\label{eq:ss_gen}
\end{subequations}

The $2\times 2$ dimension frequency-domain admittance model of a synchronous generator can be found
as:
\begin{align}
Y_{\rm gen} = -C(sI-A)^{-1}B-D
\label{eq:gen_admittance}
\end{align}
It can be seen that $(sI-A)^{-1}$ and $D$ will keep the same regardless of reference frames. On the
other hand, $C$ and $B$ are influenced by the reference frame assumption, since the generator's
$\delta$ will be different when the generator's terminal voltage is assumed as the reference
voltage or when a different reference voltage is assumed.
%

It is to be noted that the similar derivation was carried out in \cite{chevalier2018using} for
forced oscillation location research. Ref. \cite{chevalier2018using} presents not only the 2nd
order model-base admittance but also a 6th-order subtransient model-based admittance. In Section VI
of the present paper, torsional dynamics is further included in the generator admittance for a more
comprehensive study.

With the admittance model of a classical synchronous generator, two case studies are carried out
to compute eigenvalues based on the network admittance matrix. 

\paragraph{\bf Example 1: SMIB system}
We will use the well-known single-machine infinite bus (SMIB) system for illustration of
admittance-based linear analysis. The generator is connected through a line (pure reactance $jX_L$)
to the infinite bus ($V_\infty\phase{0^{\circ}}$). If we consider a synchronous generator
represented by an internal voltage $E\phase{\delta}$ behind a transient reactance $jX_g$ has only
second-order swing dynamics, the linear system transfer function from $\Delta P_m$ to the rotor
angle $\Delta \delta$ is well known and is listed as follows.
\begin{align}
\frac{\Delta \delta}{\Delta P_m } = \frac{1}{Ms^2 + Ds +T},
\end{align}
where $M = 2H/\omega_0$, $D= D_1/\omega_0$, and $T = \frac{EV_\infty}{\tilde{X}}\cos\delta$,
$\tilde{X}$ is the total reactance including line and generator's transient reactances. The
system's characteristic polynomial is \begin{equation} Ms^2 + Ds +T. \label{eq:SMIB}\end{equation}

Using the admittance model derived in \eqref{eq:gen_admittance}, at the terminal bus, the total
admittance is
\begin{align}
Y = Y_{\rm gen} + \begin{bmatrix} 0 & \frac{1}{X_L} \\ -\frac{1}{X_L} & 0 \end{bmatrix}
\end{align}
The numerator of $Y$ is found: \begin{equation} \tilde{X}^3\omega_0( M s^2 + D s +
T).\label{eq:SMIB_Y}\end{equation}

It can be seen that the two polynomials \eqref{eq:SMIB} and \eqref{eq:SMIB_Y} share the same roots.

\paragraph{\bf Examples 2-3: Admittance-based Eigenvalue Analysis for Power Grids}

\begin{figure}[!ht]
\centering
\includegraphics[width=3.0in]{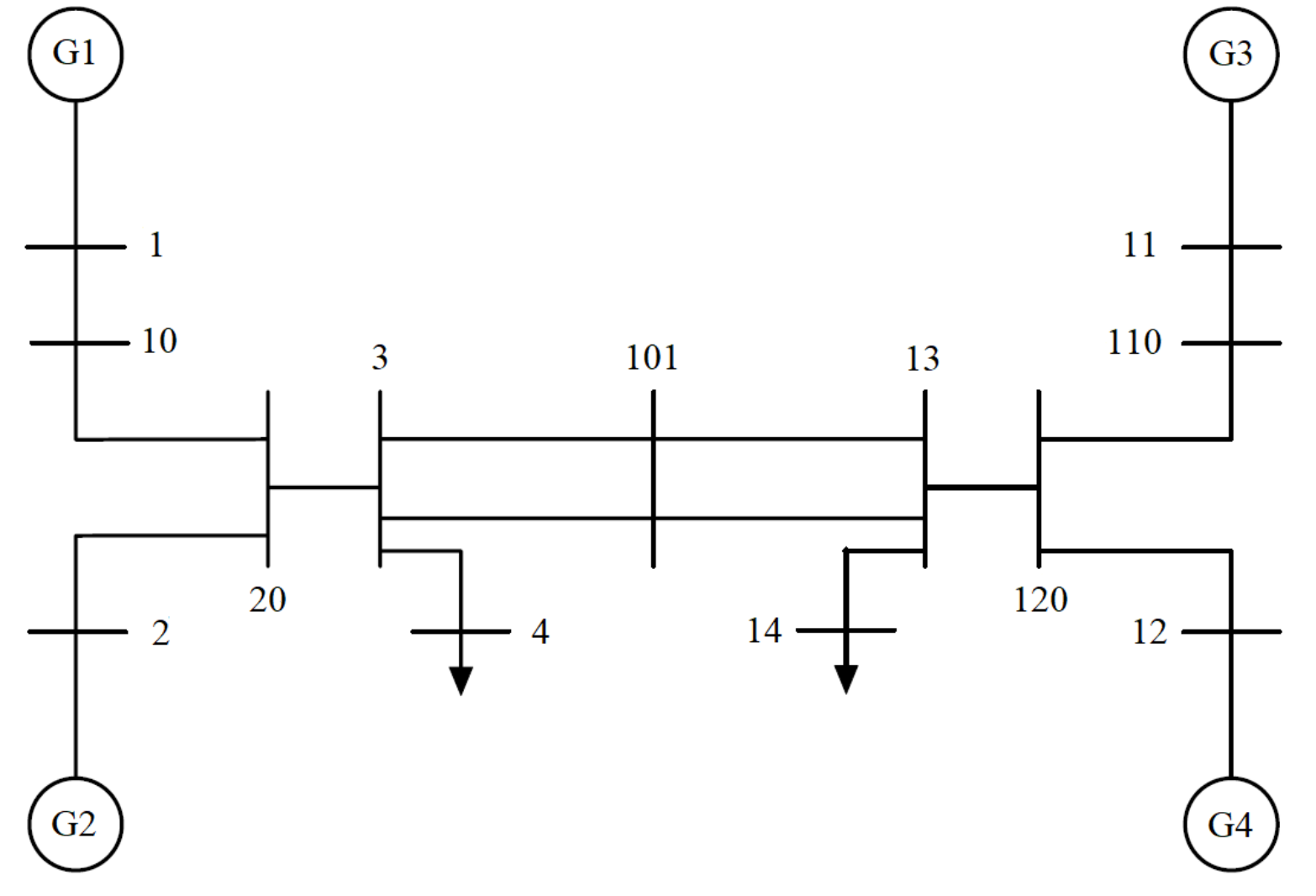}\vspace{-0.1in}
\caption{  Two-area four-machine power grid in PST.}
\label{powergrid1}
\end{figure}

\begin{figure}[!ht]
\vspace{-0.15in}
\centering
\includegraphics[width=3.0in]{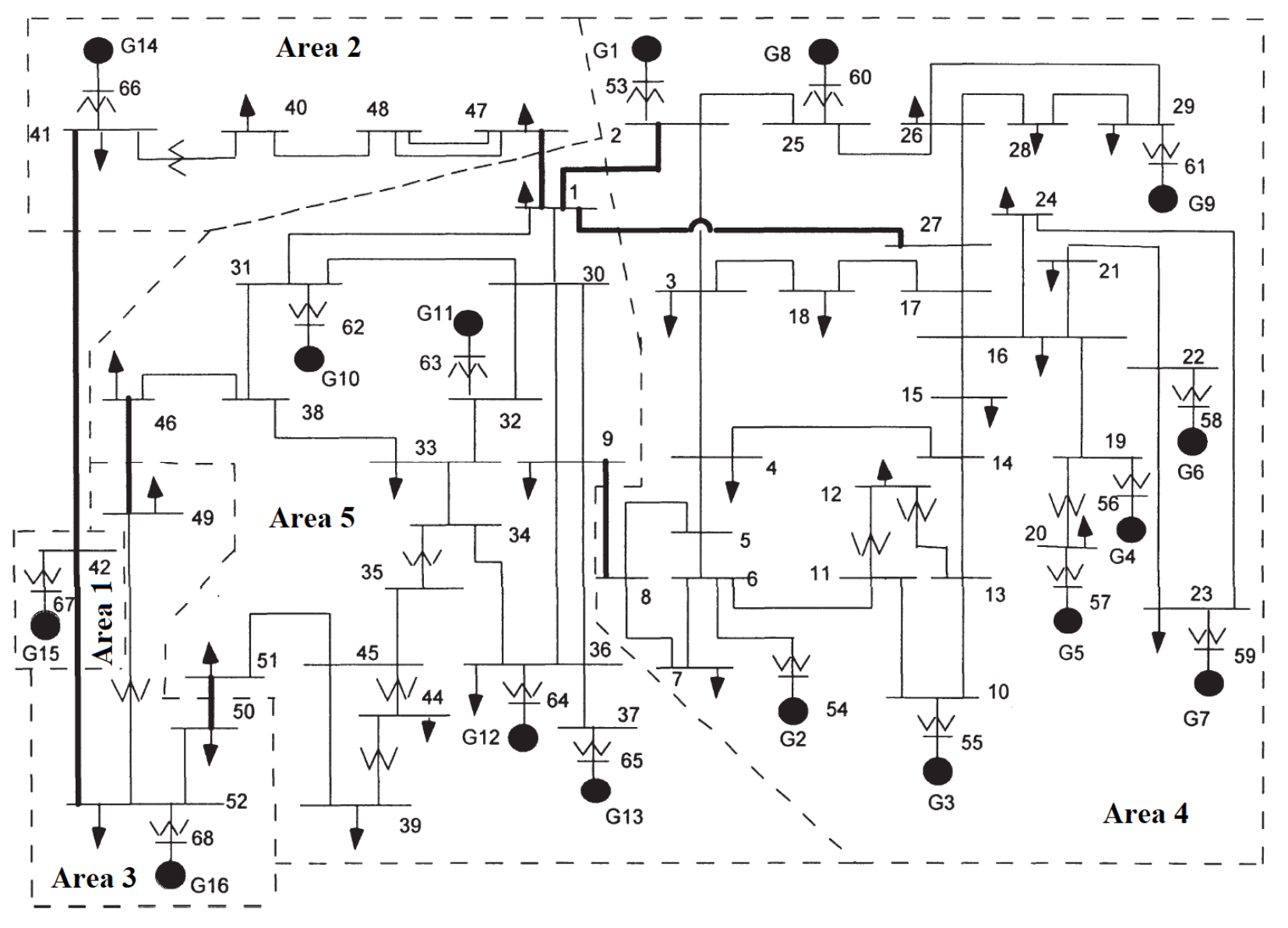}\vspace{-0.1in}
\caption{  16-machine 68-bus power grid in PST.}
\label{powergrid2}
\end{figure}

 We next demonstrate the exact match of the eigenvalue computing results of the
two-area four-machine system (Fig. \ref{powergrid1}) and the 16-machine 68-bus system (Fig.
\ref{powergrid2}) from the proposed modular analysis tool versus those from PST package. To the
authors' best knowledge, this type of validation has not been conducted before.

Classical models are assumed for all synchronous generators. Loads are assumed as constant
impedances. The 2-area 4-machine system has 8 eigenvalues while the 16-machine system has 32
examples. PST uses numerical perturbation to extract the linearized system and give eigenvalues.

In the modular approach, first, power flow is conducted. The terminal voltage phasors of the
generator buses and rotor angles of generators are found.  The information is used to compute the
$A$, $B$, $C$, $D$ matrices presented in \eqref{eq:ss_gen} and further to obtain the four
generators' admittances. For the 2-area 4-machine system, Kron reduction is carried out on the
13-bus power network to have only four generator buses left to define the passive network. Each
generator's admittance is assembled into the 4-node passive network admittance matrix ($8\times
8$). Determinant of the total admittance is then obtained and the roots of its numerator are
computed. For the 16-machine 68-bus system, 16 generator nodes are left for admittance matrix
assembling. The total admittance matrix has $32\times 32$ dimensional in the $dq$-frame.

\begin{figure}[!ht]
\centering
\begin{subfigure}{2.5in}
\centering
\includegraphics[width=2.5in]{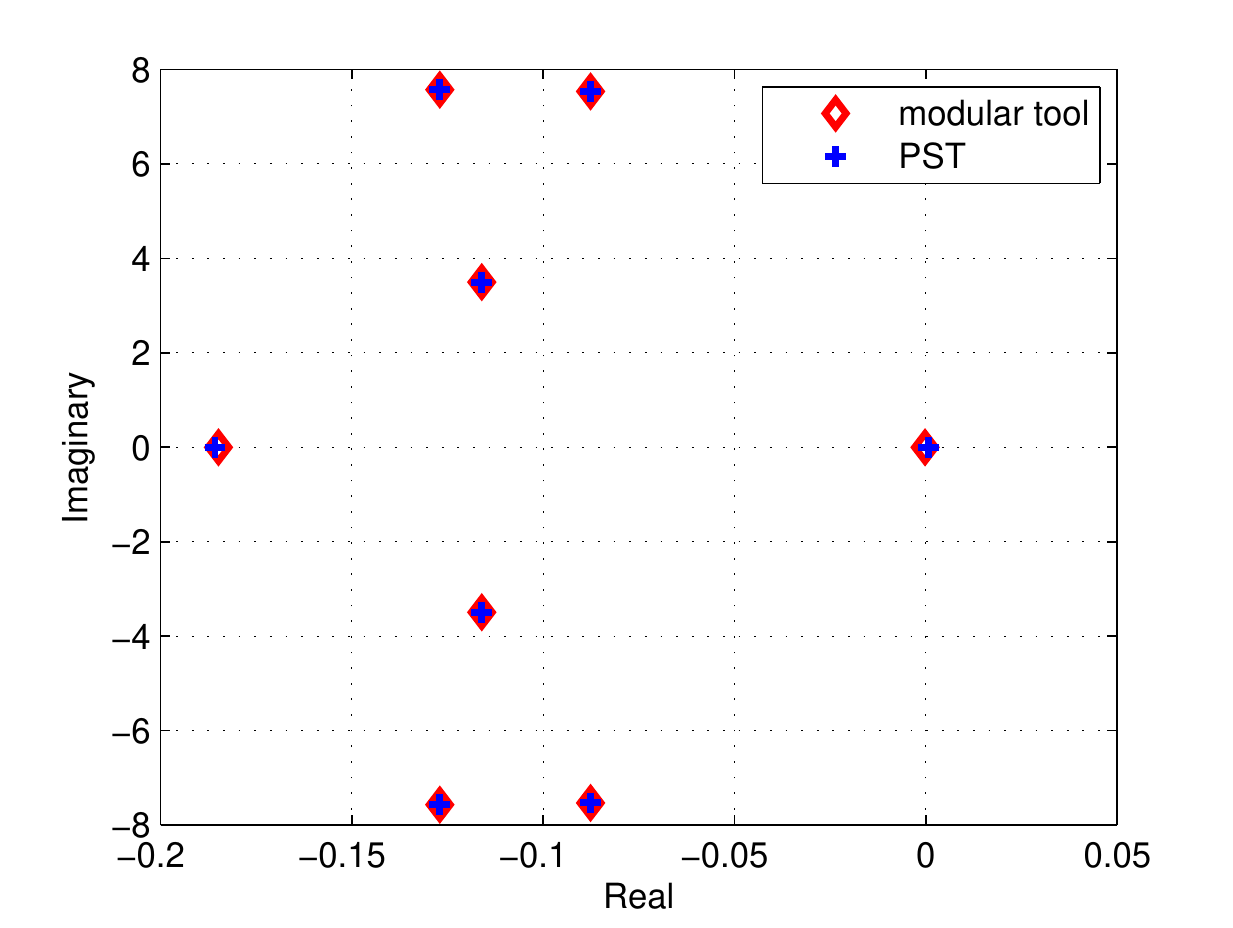}
\caption{  \label{Fig:pst_2a4m}}
\end{subfigure}
\begin{subfigure}{2.5in}
\centering
\includegraphics[width=2.5in]{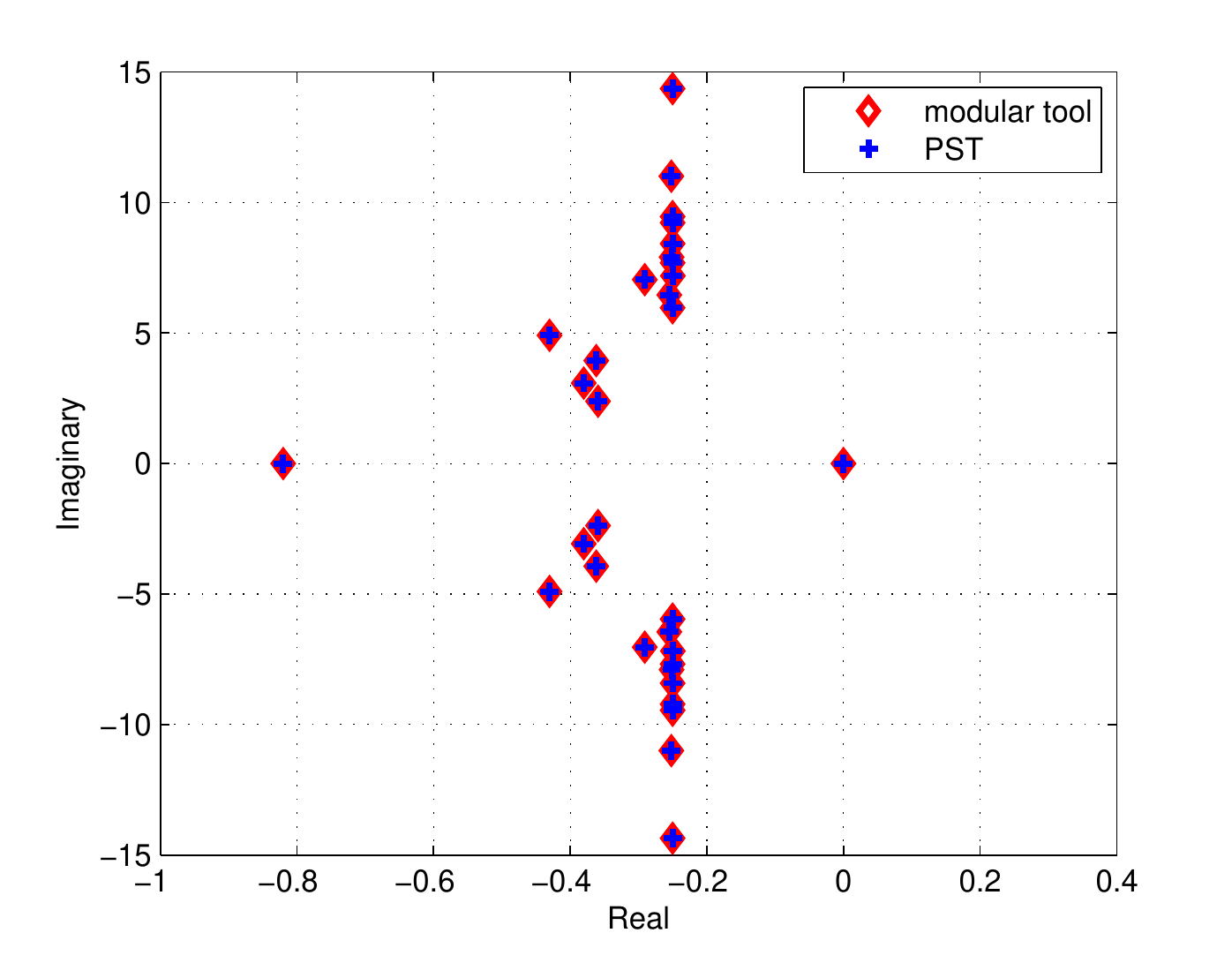}
\caption{  \label{Fig:pst_16m} }
\end{subfigure}
\caption{Comparison of eigenvalues obtained from PST and the modular tool. (a) 2-area 4-machine system. (b) 16-machine 68-bus system.}
\end{figure}

The two sets of the eigenvalues obtained from PST and the proposed modular tool are compared in
Figs. \ref{Fig:pst_2a4m} and \ref{Fig:pst_16m}. It can be seen that the results are exactly
matched.


\begin{figure*}[!ht]
\vspace{-0.1in}
\begin{subfigure}[b]{4.3in}
\centering
\includegraphics[width=4.3in]{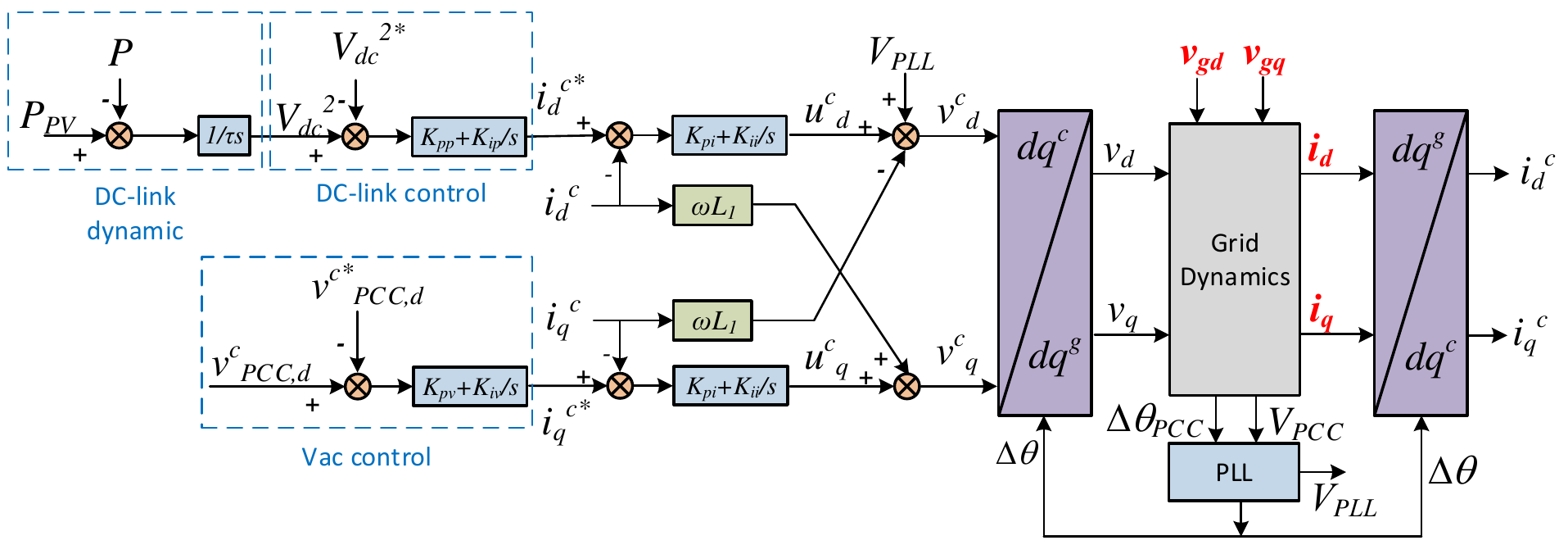}
\caption{\label{fig:blocka}}
\end{subfigure}
\begin{subfigure}[b]{2.5in}
\centering
\includegraphics[width=2.7in]{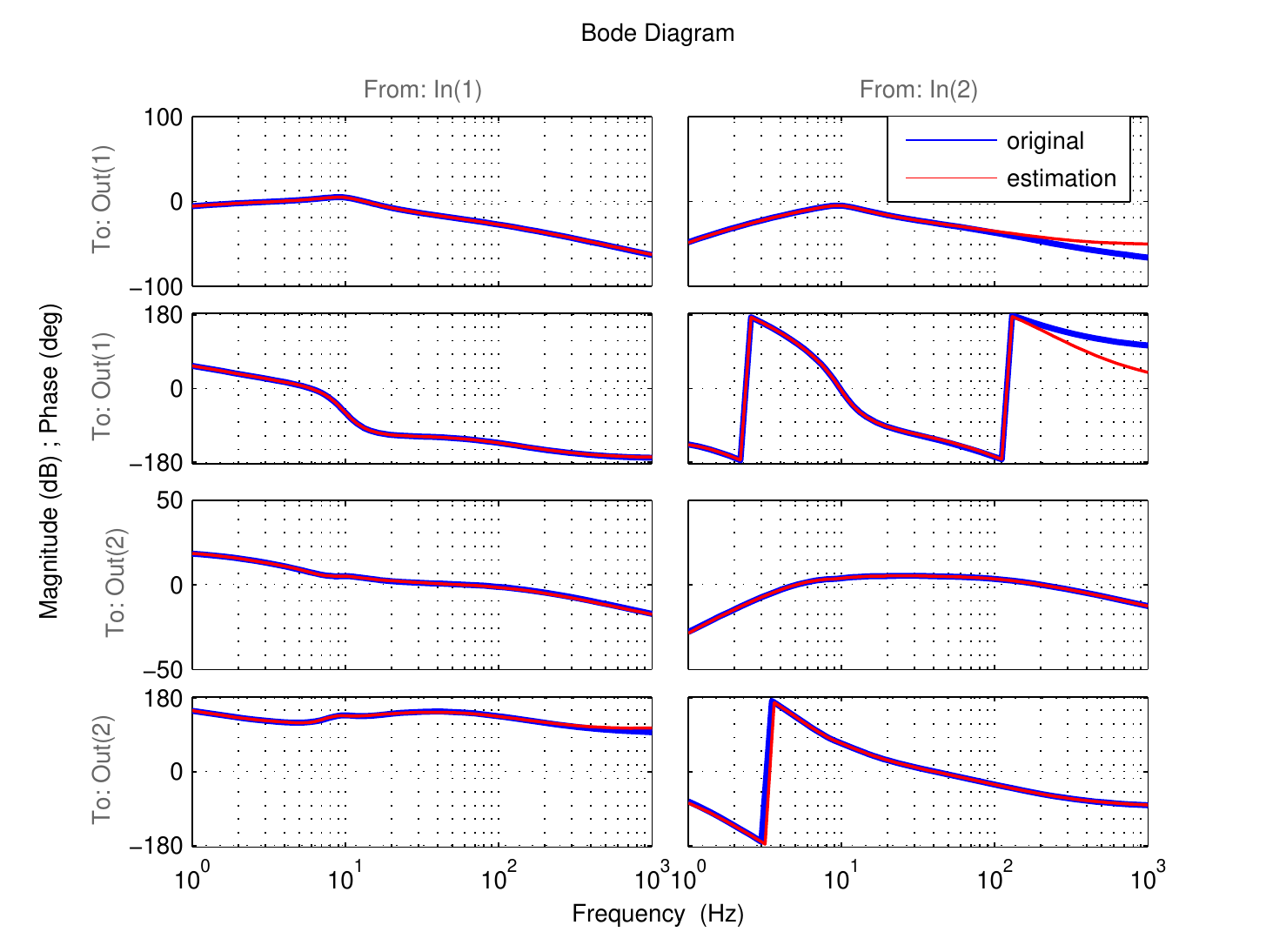}\vspace{-0.1in}
\caption{\label{fig:blockb}}\vspace{-0.10in}
\end{subfigure}
\caption{  Block diagram of the analytical model and its admittance frequency-domain responses. Parameters of the test case in pu: $(R_g, X_g): (0.001, 0.1)$, $(R_L, X_L):(0.003, 0.15)$, inner current PI controller: $(K_{pi}, K_{ii}):(0.3, 5)$,
outer loop PI parameters: $(1, 100)$, second-order PLL: $(K_{p,{\rm PLL}}, K_{i,{\rm PLL}})=(60,1400)$. Dc-link capacitor time constant $\tau$: $0.0272$ s.}
\label{Fig:block}
\vspace{-0.15in}
\end{figure*}

\subsubsection{\bf Nonlinear Model-based Approach}
\begin{figure}[!ht]
\vspace{-0.15in}
\centering
\includegraphics[width=1.0\linewidth]{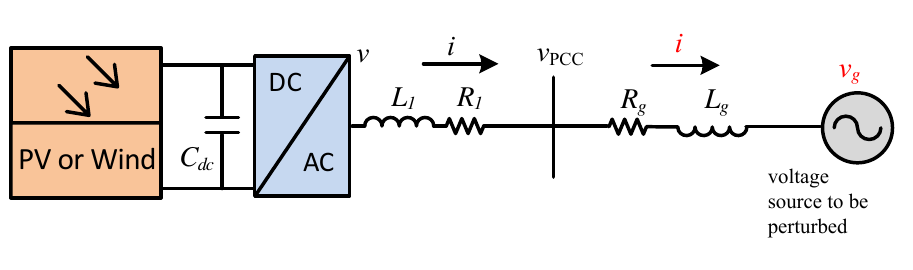}\vspace{-0.15in}
\caption{  Grid-connected VSC circuit diagram.}\vspace{-0.10in}
\label{Fig.Analytical_PV}
\end{figure}

 {\bf Example 4:} We use an example to illustrate a more efficient approach: nonlinear model-based
approach. This approach is used to obtain the $dq$-frame admittance of a VSC from the nonlinear
model. The grid-following VSC is assumed to achieve two control functions: regulate the dc-link
voltage and regulate the PCC voltage.  The admittance of the VSC viewed from the PCC bus is
desired. To find the admittance, the integration system is constructed to have the PCC bus directly
connected to the grid voltage source. Since the VSC regulates the PCC voltage, direct connection of
a voltage source at PCC bus is not desirable since it yields the PCC voltage not controllable.
Rather, a very small $R_g$ and $L_g$ will be inserted between the grid voltage source and the PCC
bus. The topology is shown  in Fig. \ref{Fig.Analytical_PV}.

The analytical model of the system is constructed in the $dq$-frame. This model has been used for
stability analysis related to VSC in weak grids \cite{fan2018modeling, fan2018wind, li2019wind,
li2018stability}. The full details of the modeling blocks are shown in Fig. \ref{Fig:block}(a).
Using numerical perturbation (e.g., MATLAB command \texttt{linmod}), lineraized model at an
operating condition can be found. An input/output linearized model is found with the $dq$-axis
voltages as input and the $dq$-axis currents as output, shown in the following. 
\begin{align}
\begin{bmatrix}
i_{d}(s) \\
i_{q}(s)\\
\end{bmatrix}
=-
\underbrace{
\begin{bmatrix}
Y_{dd}(s)& Y_{dq}(s) \\
Y_{qd}(s) & Y_{qq}(s) \\
\end{bmatrix}}_{Y_{{\rm vsc},dq}}
\begin{bmatrix}
v_{gd}(s) \\
v_{gq}(s)\\
\end{bmatrix}
\label{Yddqq}
\end{align}
Fig. \ref{Fig:block}(b) (``original'') presents the frequency-domain responses of the admittance
matrix.


{\bf Remarks:} This approach to find admittance model can be adopted for any component, including a
synchronous generator should its analytical model is available. Through terminal voltage
perturbation, a $dq$-frame admittance may be found.




\subsection{Black-Box Approach: Finding impedance through time-domain responses}
\label{section_blackbox}
From time-domain responses to find an SISO model or a transfer function has been investigated in
1990s for control design  \cite{sanchez1997computation}. To obtain the plant model from the exciter
input to the speed, impulse signals are injected and the speed responses are recorded. The data of
the speed is fed into Eigensystem Realization Algorithm (ERA) and the transfer function is
obtained. Other similar methods that lead to eigenvalue identification are Prony analysis and
Matrix Pencil \cite{crow2012identification}. We will adopt ERA to find VSC's $dq$-frame admittance
matrix using time-domain responses. To have a clear picture of ERA, the algorithm is first
reviewed. Challenges of applying ERA for admittance matrix identification is then laid out. A
modified ERA is demonstrated to tackle the challenge.

\subsubsection{ERA for measurement data generated by single event} ERA was proposed by Juang and Pappa in 1985
\cite{juang1985eigensystem} to find a linear model from measurement data in the field of aerospace.
ERA assumes that the dynamic response is due to impulse input.  Consider a Linear-Time Invariant
(LTI) system in discrete domain as the following:
\begin{equation}
{x}_{k+1}=Ax_k+Bu_k, y_k=Cx_k+Du_k\label{eq:27}
\end{equation}
where $y \in \mathbb{R}^{K\times 1}$ is defined as the output column vector of the system with $K$
output channels, $A \in \mathbb{R}^{n\times n}$, $B \in \mathbb{R}^{n\times 1}$, $C \in
\mathbb{R}^{K\times n}$, and $D \in \mathbb{R}^{K\times 1}$ are system matrices. Assuming $x_0=0$,
the system response due to an impulse input ($u_0 = 1$, $u_k=0, k>0$) can be found as follows.
\begin{align}
&x_0 = 0, x_1 = B, x_2 = AB, \cdots, x_k = A^{k-1}B \notag\\
 & y_0 = D,  y_1 = CB, y_2 = CAB, \cdots, y_k = CA^{k-1}B\notag
\end{align}
Two shifted Hankel matrices are formed as follows.
\begin{align}
H_1 &=\begin{bmatrix}
y_1 & y_2 & \cdots & y_{L} \\
y_2 & y_3 & \cdots & y_{L+1} \\
\vdots & \vdots & \ddots & \vdots \\
y_{N-L+1} & y_{N-L+2} & \cdots & y_{N}
\end{bmatrix}_{K(N-L+1)\times L} \notag \\
H_2 &=\begin{bmatrix}
y_2 & y_3 & \cdots & y_{L+1} \\
y_3 & y_4 & \cdots & y_{L+2} \\
\vdots & \vdots & \ddots & \vdots \\
y_{N-L+2} & y_{N-L+3} & \cdots & y_{N+1}
\end{bmatrix}_{K(N-L+1)\times L}
\notag
\end{align}

It can be seen that the Hankel matrices can be decomposed as follows.
\begin{align}
H_1 &=\begin{bmatrix}
CB & CAB & \cdots & CA^{L-1}B \\
CAB & CA^2B & \cdots & CA^LB \\
\vdots & \vdots & \ddots & \vdots \\
CA^{N-L}B & CA^{N-L+1}B & \cdots & CA^{N-1}B
\end{bmatrix}
\notag \\
&=\underbrace{\begin{bmatrix}
C\\
CA\\
\vdots\\
CA^{N-L}
\end{bmatrix}}_{\mathcal{O}}
\underbrace{\begin{bmatrix}
B & AB & \cdots & A^{L-1}B
\end{bmatrix}}_{\mathcal{C}}
=\mathcal{O}\mathcal{C}
\\
H_{2}&=\mathcal{O}A\mathcal{C} \label{eq:ERA_H2}
\end{align}
where $\mathcal{O}$ is the observability matrix and $\mathcal{C}$ is the controllability matrix.
Note that the two matrices are of the following dimensions:
\begin{align*}
\mathcal{O}  \in \mathbb{R}^{K(N-L+1) \times n}, \>\>
\mathcal{C}\in \mathbb{R}^{ n \times L}
\end{align*}
ERA employs singular value decomposition (SVD) and rank reduction to find two matrices to realize
$\mathcal{O}$ and $\mathcal{C}$. First, SVD is conducted for $H_1$ and the resulting matrices are
listed as follows with their dimensions notated.
\begin{equation}
\begin{aligned}
H_{1} &=USV^{T}, \\
\text{where  } &U\in \mathbb{R}^{K(N-L+1) \times K(N-L+1)}, \\
&S\in \mathbb{R}^{K(N-L+1)\times L },\>\>
V\in \mathbb{R}^{L \times L}
\end{aligned}
\end{equation}
Only $n$ components of $\text{diag}(S)$ will be kept to construct the reduced-rank Hankel matrix
$H'_1$.
\begin{equation}
\begin{aligned}
H'_{1} &=\underbrace{U(:, 1: n)}_{U'} \underbrace{S(1:n, 1:n)}_{S'} (\underbrace{V(:, 1:n)}_{V'})^{T} \\
&U'\in \mathbb{R}^{K(N-L+1) \times n },
S'\in \mathbb{R}^{n\times n }, \>V'\in \mathbb{R}^{L \times n}
\end{aligned}
\end{equation}

Similarly, rank reduction may also be applied to $H_2$ to have a low-rank Hankel matrix $H_2'$. 
From the reduced-rank Hankel matrix, the observability and controllability matrices with correct
dimensions can be realized.
\begin{align}
\mathcal{O} = U'S'^{\frac{1}{2}}, \>\>\>
\mathcal{C} = {S'}^{\frac{1}{2}}(V')^{T}
\end{align}

Thus, the system matrix $A$ can be realized through the use of \eqref{eq:ERA_H2}. Moreover, $B$ and
$C$ can be found from the controllability matrix and the observability matrix. $D$ is the data at
the initial time.
\begin{align}
A &= S'^{-\frac{1}{2}}U'^T H'_2 V'{S'}^{-\frac{1}{2}},\>\>
B  =\mathcal{C}(:,1), \notag\\
C & = \mathcal{O}(1:K,:), \>\>
D  = y_0.
\end{align}
The entire state-space model is known. Hence, the eigenvalues of the system and each signal's
transfer function are all known.

\subsubsection{\bf Example 5: Identification of Admittance for a VSC}
The $dq$-frame admittance of the  grid-connected VSC (shown in Fig. \ref{Fig.Analytical_PV}) will
be
identified. 
This 9th order $dq$-frame based model,
shown in Fig. \ref{fig:blocka} is used to produce data. Small-signal model can be extracted from
the analytical model at an operating condition through numerical perturbation. 
This model can serve as the
benchmark for the measurement-based admittance.

The foremost challenge is how to create data that is suitable for admittance identification. In
harmonic injection method, sinusoidal perturbation is used. On the other hand, for time-series
data, Laplace transform of the step response of a system is associated with the product of the
transfer function of the system and $1/s$. Thus, step changes may be used as perturbation. This
requires the perturbed variables and measurements be constant at steady-state. Hence, $dq$-frame
models will be suitable to generate step responses.

\begin{figure}[!ht]
\vspace{-0.15in}
\centering
\includegraphics[width=0.78\linewidth, height=2.68in]{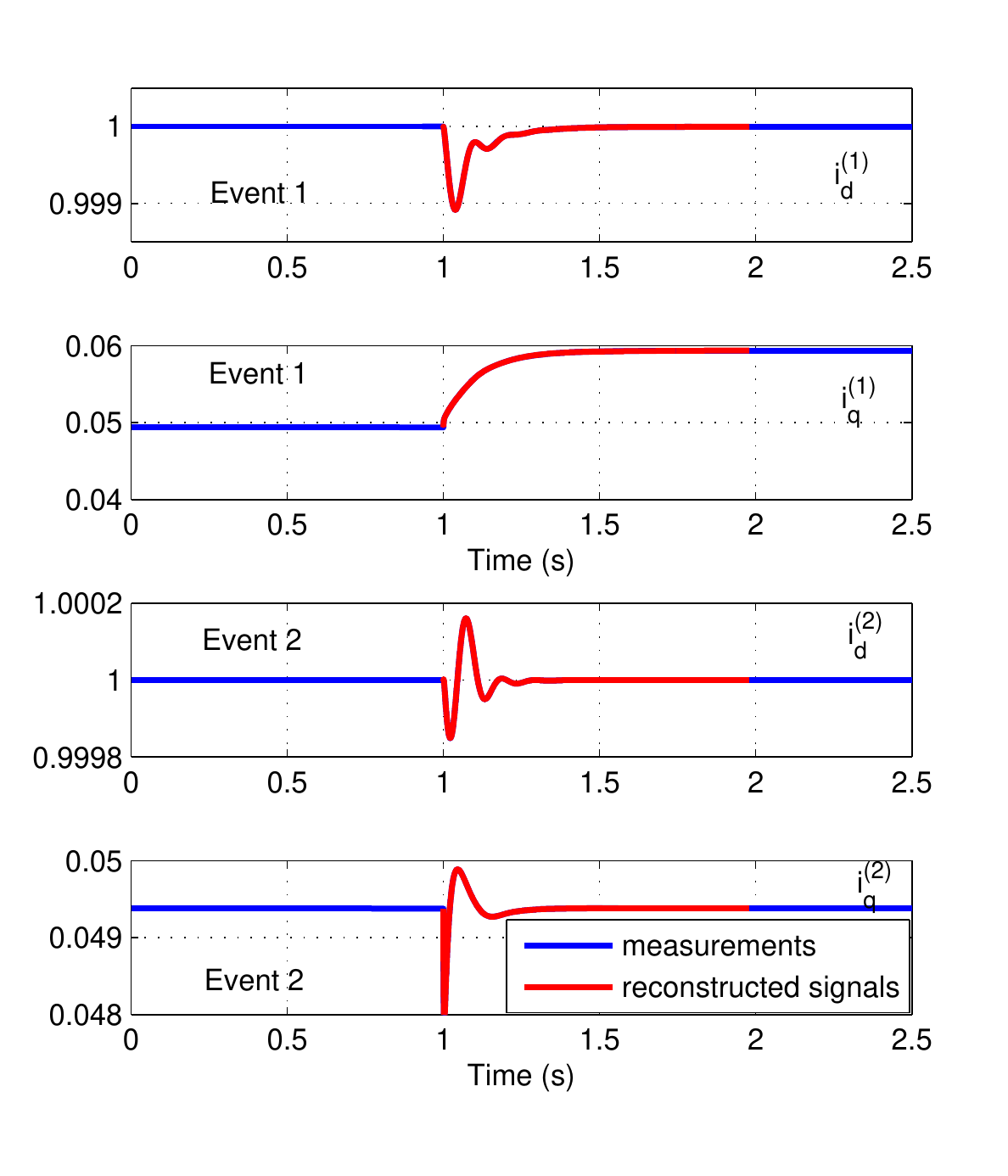}\vspace{-0.2in}
\caption{  Dynamic responses of $i_d$ and $i_q$ for two events. Event 1: $v_{gd}$ step change of $0.001$ pu
 at $t=1$ s. Event 2: $v_{gq}$ step change of $0.001$ pu at $t=1$ s. }\vspace{-0.10in}
\label{fig:measurement}
\end{figure}

A step change with 0.001 pu size will be applied to $v_{gd}$. Line currents are measured and
notated as $i^{(1)}_d$ and $i^{(1)}_q$. Another step change with 0.001 pu size is applied to
$v_{gq}$ and the line currents are notated as $i^{(2)}_d$ and $i^{(2)}_q$. The step response data
are presented in Fig. \ref{fig:measurement}.

Data in the time frame from 1 second to 1.98 is used for analysis. Sampling frequency is 2500 Hz.
Since ERA assumes the initial state variables are zeros, the data are processed to have the initial
steady-state values taken off. Proper scaling is applied to signals to have similar degree of
variation. For the four signals, the scales are $1000$, $1$, $100$, and $10$. The two sets of
scaled event data will be used.



\paragraph{ERA for MIMO systems}
A distinct feature of the $dq$-admittance model is that it is a two-input two-output system. Since
the dynamic event measurements are generated from the same dynamic system, the estimated system
dynamic matrix $A$ is expected to be the same. If the two event data are treated separately, it is
possible that the resulting $A$ matrix is different in two cases. To solve this challenge, a
modified ERA that can handle multiple event data is worked out by the authors.

The two event data are generated based on $dq$-axis voltage perturbation, respectively. Each event
corresponds to a single-input two-output system. For the two events, the inputs are different,
while the outputs are the same. Hence, the corresponding The two linear systems share the same $A$
and $C$ matrices. The $B$ and $D$ matrices for the two systems are different.
\begin{align}
\begin{aligned}
\text{Event 1:}\>\> {x}_{k+1}&=Ax_k+ B^{(1)} u^{(1)}_k\\
y^{(1)}_k& = C x_k +D^{(1)} u^{(1)}_k\\
\end{aligned}\\
\begin{aligned}
\text{Event 2:}\>\>x_{k+1}&=Ax_k+ B^{(2)} u^{(2)}_k\\
y^{(2)}_k& = C x_k +D^{(2)} u^{(2)}_k
\end{aligned}
\end{align}
where superscript $(i)$ notates $i$th event related system.

%

For each event data, two shifted Hankel matrices may be formed. We already know that they can be expressed as the multiplication of the observability matrix, $A$, and the controllability matrix. The observability matrices for the two events will be the same, while the controllability matrices for the two events will be different.
\begin{align}
\begin{aligned}
H^{(1)}_1 &= \mathcal{O}\mathcal{C}^{(1)}, \>\>H^{(2)}_1 = \mathcal{O}A\mathcal{C}^{(2)}, \\
H^{(1)}_{2}&=\mathcal{O}\mathcal{C}^{(2)}, \>\>H^{(2)}_{2}=\mathcal{O}A\mathcal{C}^{(2)}
\end{aligned}
\end{align}

The first two Hankel matrices may form a new Hankel matrix while the latter two Hankel matrices may form another shifted new Hankel matrix. Their relationship can be found as follows.

\begin{align}
\begin{aligned}
H_1 &= \begin{bmatrix} H^{(1)}_1 & H^{(2)}_1\end{bmatrix} = \mathcal{O}\begin{bmatrix} \mathcal{C}^{(1)} &\mathcal{C}^{(2)} \end{bmatrix}\\
H_2 &= \begin{bmatrix} H^{(1)}_2 & H^{(2)}_2\end{bmatrix} = \mathcal{O}A \begin{bmatrix} \mathcal{C}^{(1)} &\mathcal{C}^{(2)} \end{bmatrix}
\end{aligned}
\end{align}
The two Hankel matrices formed by two event data now assume the same relationship of the two formed
by single event data. The same procedure that is used to find $A$ matrix can be used. Further, $B$
and $C$ matrices can be found.

\paragraph{Study results}
The two sets of the data are fed into a modified ERA that can handle data from multiple events. The
estimated system is assumed to have an order of $10$. This order is chosen by considering the 9th
order system and the step response perturbation.

The system matrix $A$ is first computed. The eigenvalues of the continuous system are estimated and
the residuals for each signal can be computed. With the eigenvalue and residuals found, measurement
data can be reconstructed and the transfer function can be found for each signal. Fig.
\ref{fig:measurement} presents the reconstructed signals with their initial steady-state values
added. The reconstructed signals have a close  to 100\% match with the original data.

\begin{figure}[!ht]
\vspace{-0.10in}
\begin{subfigure}[b]{1.72in}
\centering
\includegraphics[width=1.72in]{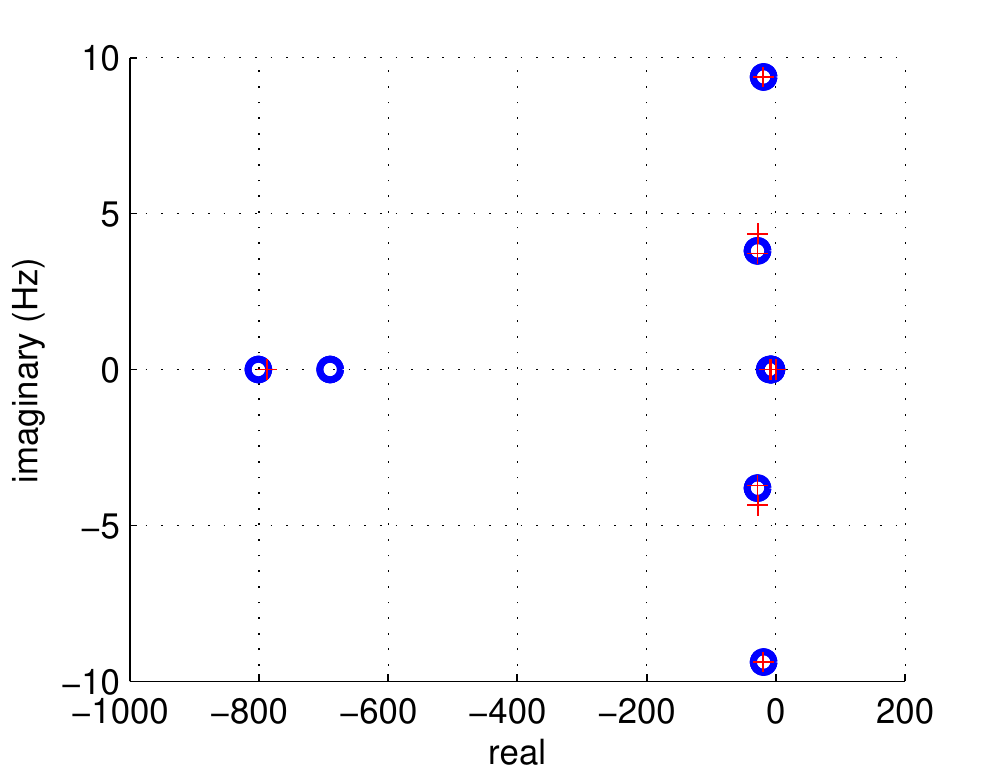}
\end{subfigure}
\begin{subfigure}[b]{1.72in}
\centering
\includegraphics[width=1.72in]{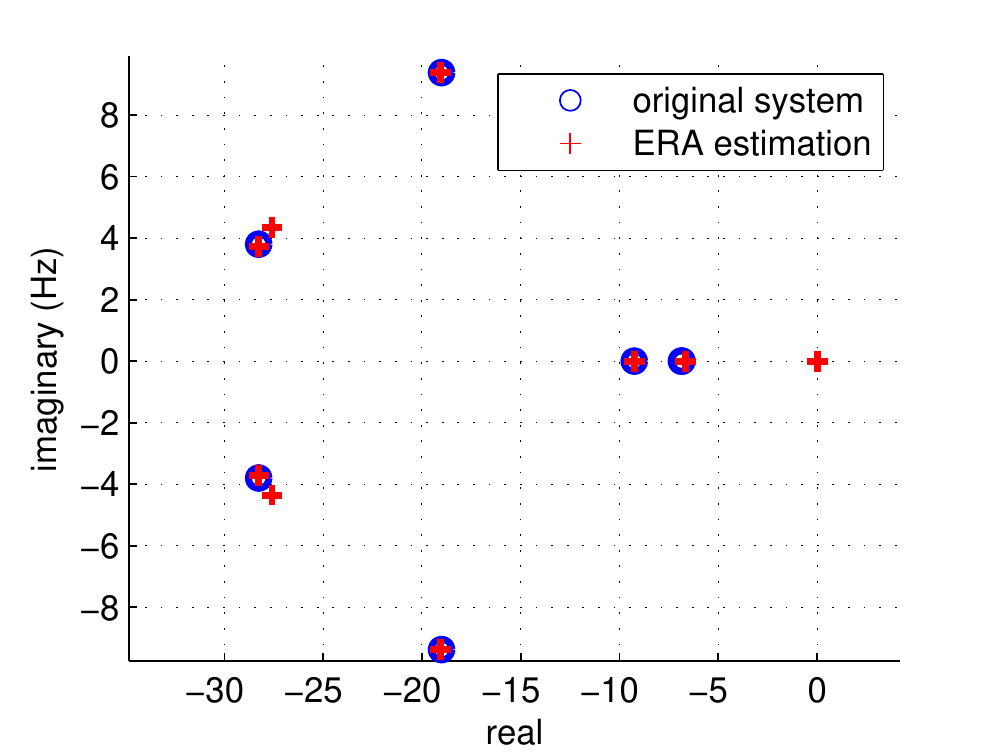}
\end{subfigure}
\caption{  Actual eigenvalues of the 9th order system and the estimation from current signals. The right figure is the zoom-in of the left figure.  }\vspace{-0.10in}
\label{Fig:eig}
\end{figure}

Fig. \ref{Fig:eig} presents the estimated eigenvalues versus the eigenvalues of the original system
in the complex plane. 
It can be seen that the identified eigenvalues align
very well with the original eigenvalues. The estimated eigenvalue at the original point is due to
the step input which introduces an ``s'' in the denominator of the current transfer functions.

For each signal, its estimated Laplace domain expression can now be found: $i^{(1)}_d(s)$,
$i^{(1)}_q(s)$, $i^{(2)}_d(s)$, $i^{(2)}_q(s)$, where superscript notates event number. The
admittance model can be found by taking into the effect of step response of $v_d$ or $v_q$.
\begin{equation}
Y= -\frac{s}{p}\begin{bmatrix}
i^{(1)}_d(s)& i^{(2)}_d(s) \\
i^{(1)}_q(s)& i^{(2)}_q(s)
\end{bmatrix}
\end{equation}
where $p$ is the size of perturbation. For this study, $p = 0.001$.

Fig. \ref{fig:blockb} presents the frequency responses of the estimated admittance versus the
original admittance. It can be seen that the estimated admittance gives excellent match in the
frequency range of 1 Hz to 100 Hz. Considering that grid dynamic studies are concerned of dynamics
in this range, the proposed approach, using step responses of 1 second data for admittance
identification, has been demonstrated as a powerful tool for device characterization.









\section{Vision 2: Assembling into A Network Admittance Matrix}
While majority of the current research on stability analysis of inverter penetrated systems
considers an entire system as two impedances, a few also adopted the notation of impedance network
or network admittance.

In \cite{liu2018oscillatory, liu2017stability}, components of a power grid, including individual
wind farms, are modeled as impedances. Topology of a network with impedances is illustrated in
\cite{liu2018oscillatory, liu2017stability}. On the other hand, multi-node network admittance
matrix assembling is not mentioned, nor used. The authors continued with circuit branch aggregation
using manual methods to aggregate series branches and shunt branches to arrive at two impedances
viewed from the terminal bus of a wind farm.

 Assembling
component admittances into an network admittance matrix of dimension $N\times N$ (where $N$ is the
number of node) is a familiar task to power system engineers. The admittance matrix formed for
power flow analysis is essentially a steady-state admittance matrix in phasor domain.

In \cite{xu2005harmonic}, $N\times N$ complex network admittance matrices or $2N\times 2N$ network
admittance matrices (with real and imaginary part separated) are employed for harmonics analysis.
In this case, frequency-domain network admittance matrix is formed, considering passive components.
In \cite{semlyen1999sdomain}, $s$-domain admittance further includes sources.


In this paper, we give detailed formulation procedure to assemble components into a $dq$-frame
$2N_s\times 2N_s$ (where $N_s$ is the number of sources) network admittance matrix. $dq$-frame
admittance is preferred since $dq$-frame based modeling has been a tradition for synchronous
generators.

Consider a $N$-node power network with its passive elements, e.g., line, transformers and constant
impedance loads, all expressed by an admittance matrix. The system can be reduced into a network
with only the source buses. This reduction can be achieved through Kron reduction.
\begin{equation}
I_s=Y_{\rm red} V_s
\end{equation}
where $I_s$ notates the vector of current injection by sources, $V_s$ notates the vector of voltage
of the source buses, and $Y_{\rm red}$ represents the reduced-size network obtained by Kron
reduction.

If every source (synchronous generators, wind, and solar) can be represented by a Norton equivalent
as a current source in parallel with an admittance ($I_{si} = I_{gi} - y_{gi}V_{si}$), then the
entire system may be represented by an admittance matrix $Y$.
\begin{equation}
I_g =\underbrace{(Y_g +Y_{\rm red})}_{Y} V_s
\label{eq1}
\end{equation}
where $Y_g$ is a diagonal matrix with each diagonal element the admittance of a source. This
assembling procedure has also been mentioned in the 1976 paper \cite{Undrill1976}.

To examine system stability, the following MIMO system is examined.
\begin{equation}
V_s =\underbrace{Y(s)^{-1}}_{Z(s)} I_g
\label{eq2}
\end{equation}
Thus, if the MIMO transfer function matrix, represented by the impedance matrix $Z(s)$ is proper,
 the system is stable. The system eigenvalues are the roots of $\det(Y(s))=0$.


Accurate assembling is necessary to lead to accurate stability analysis results. In the following,
we present two challenges, namely, initial operating condition consideration for admittance
characterization and reference frame consideration.
\subsection{Challenge 1: Accurate Initial Condition Setup}

\begin{figure}[!ht]
\centering
\includegraphics[width=2.8in]{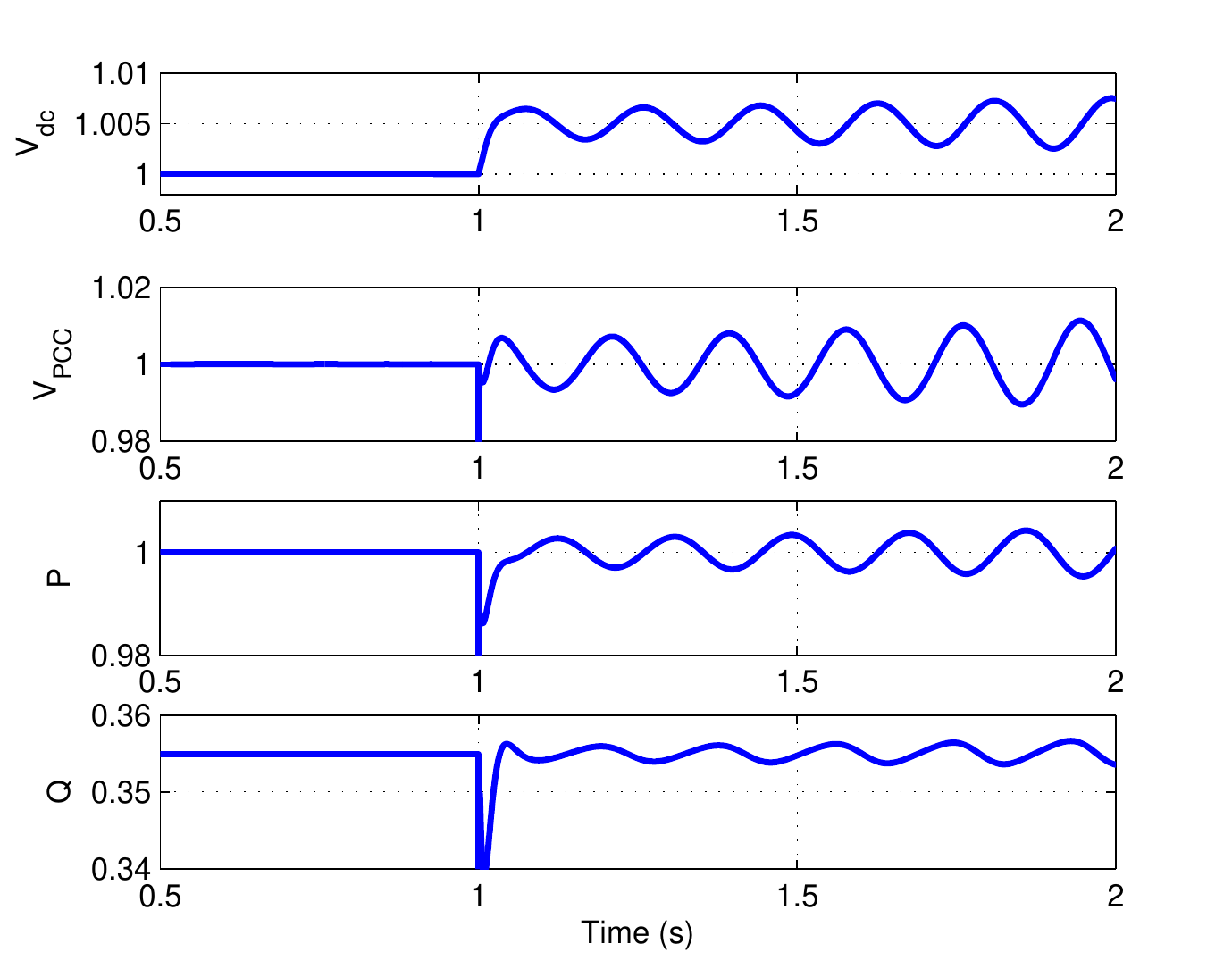}
\caption{  Dynamic responses of the VSC grid integration system subject to 0.5\% change in dc-link voltage order. $R_g = 0.08$ pu, $X_g=0.8$ pu. The system is unstable with growing 5 Hz oscillations. }
\label{Fig:VSCweakgrid_sim}
\end{figure}

We start from an example: VSC in weak grids ({\bf Example 6}). The analytical model shown in Fig.
\ref{Fig:block} has been derived in our prior work \cite{fan2018modeling,fan2018wind, li2019wind,
li2018stability} and can be used to demonstrate low-frequency oscillations. Fig. \ref{Fig:VSCweakgrid_sim} presents the system response subject to a
small step change in dc-link voltage order.

We now approach to view the system as a one-node system with two shunt admittances: $Y_{\rm VSC}$
and $Y_{\rm line}$. The node is the PCC bus. The VSC admittance can be found using the methods
presented in Section III.A or B.

For the initial operating condition ({\bf \emph{Initial condition 1}}), the power output from the
converter is assumed to be 1 pu and
the PCC voltage is 1 pu. The voltage source to be perturbed is set at $1\phase{0}$ pu. 

With the VSC admittance available, the total admittance viewed at the PCC bus is as follow:
\begin{align}
Y = Y_{\rm VSC}+ Y_{\rm line}.
\end{align}
where $Y_{\rm line} = \begin{bmatrix}R_g +sL_g & -\omega_0 L_g \\ \omega_0 L_g & R_g+sL_g
\end{bmatrix}$ and $\omega_0$ is the nominal frequency 377 rad/s.

\begin{figure}[!ht]
\centering
\begin{subfigure}{2.5in}
\includegraphics[width=2.5in]{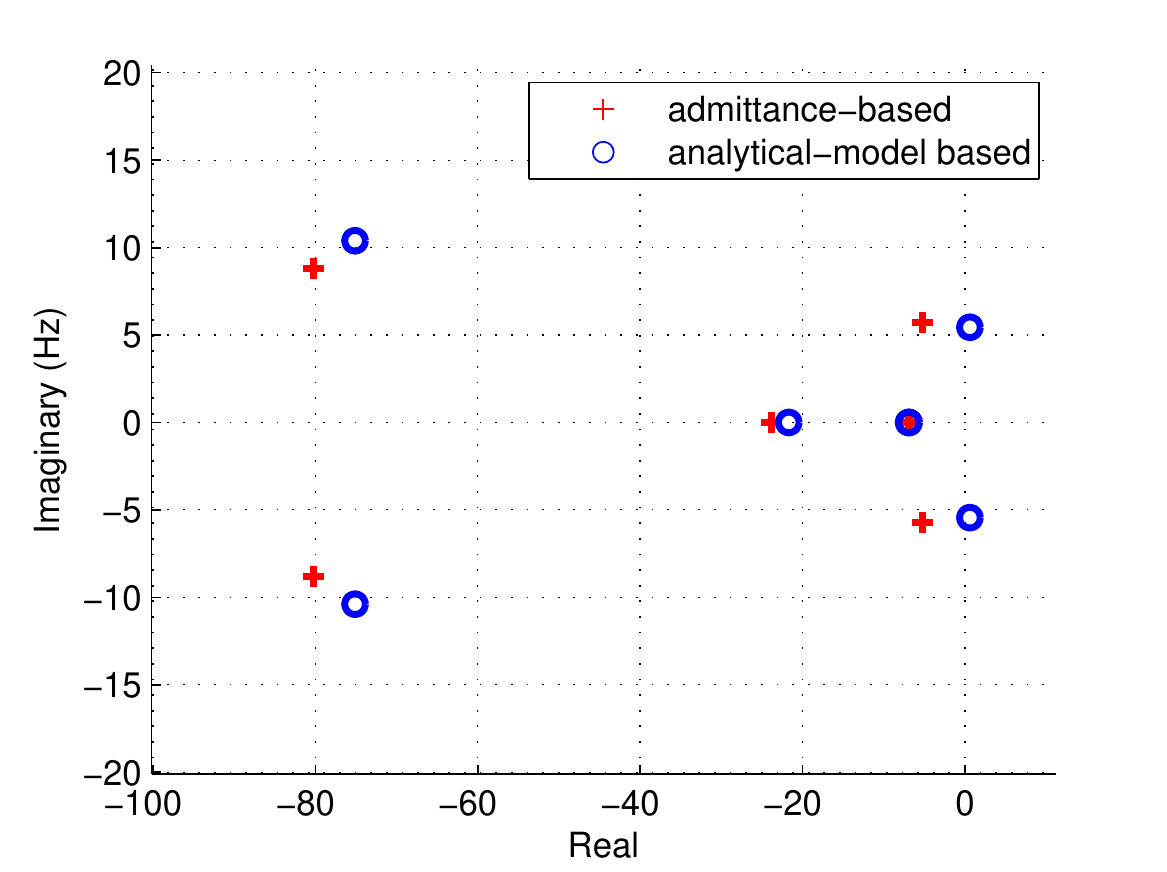}
\caption{\label{fig:VSCweakgrid1}}
\end{subfigure}
\begin{subfigure}{2.5in}
\includegraphics[width=2.5in]{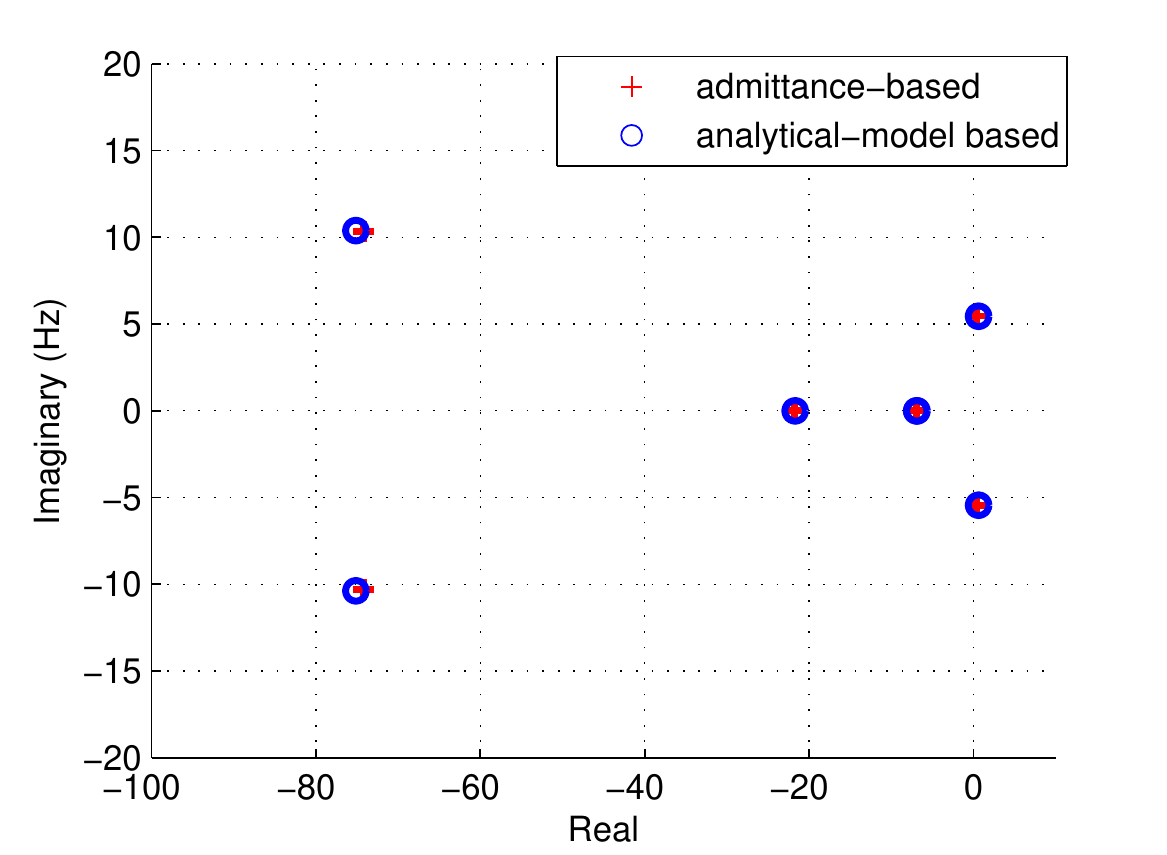}
\caption{\label{fig:VSCweakgrid2}}
\end{subfigure}
\caption{  Comparison of eigenvalues obtained using admittance-based approach and analytical model-based approach. (a) Initial condition 1.
 (b) Initial condition 2. }
\label{Fig:VSCweakgrid_eig}
\end{figure}

The system's eigenvalues are the roots of $\det(Y)$. The eigenvalues, along with the eigenvalues of
the analytical model obtained at an operating condition of $P=1$ pu, $V_{\rm PCC} =1$ pu, and
$X_g = 0.8$ pu are plotted in Fig. \ref{fig:VSCweakgrid1}.

It can be seen that the eigenvalues from the analytical model correctly indicate the system is
unstable with a pair of eigenvalues located in the RHP. On the other hand, the admittance-based
modular tool indicates that the system is stable with the dominant mode located in the LHP.

\textbf{\emph{Mitigation}} The inaccurate eigenvalues indicate that the admittance matrix obtained using Initial Condition 1 is not suitable. For both the VSC with weak grid integration testbed and
the  VSC admittance measurement testbed, the initial conditions are compared. Table
\ref{table:operating_condition} lists the operating condition for the VSC at two testbeds. It can
be observed that for the two systems, the operating condition of the VSC is very different on
reactive power generation. The measurement testbed's voltage source is then configured so that VSC
exports the same level of real or reactive power as the original testbed ({\bf \emph{Initial
Condition 2}}). The voltage source is configured based on the PCC frame. Fig.
\ref{fig:VSCweakgrid2} presents the eigenvalues of the admittance-based approach and the original
set. They show exact match.
\begin{table}[ht!]
\centering
\caption{Operating condition comparison}
\label{table:operating_condition}
\begin{tabular}{ccc}
\hline\hline
 &Original Testbed & Measurement Testbed \\
 \hline
$P$ (pu)  & 1 & 1 \\
$V_{\rm PCC}$  (pu) & 1 & 1 \\
$Q$ (pu) & -0.355 & -0.09\\
$\Delta \theta_{\rm PCC}$ & $50.5^\circ$ & $0.57^\circ$\\
\hline
\hline
\end{tabular}
\end{table}


\begin{figure}[!ht]
\begin{subfigure}{3.5in}
\centering
\includegraphics[width=3.5in]{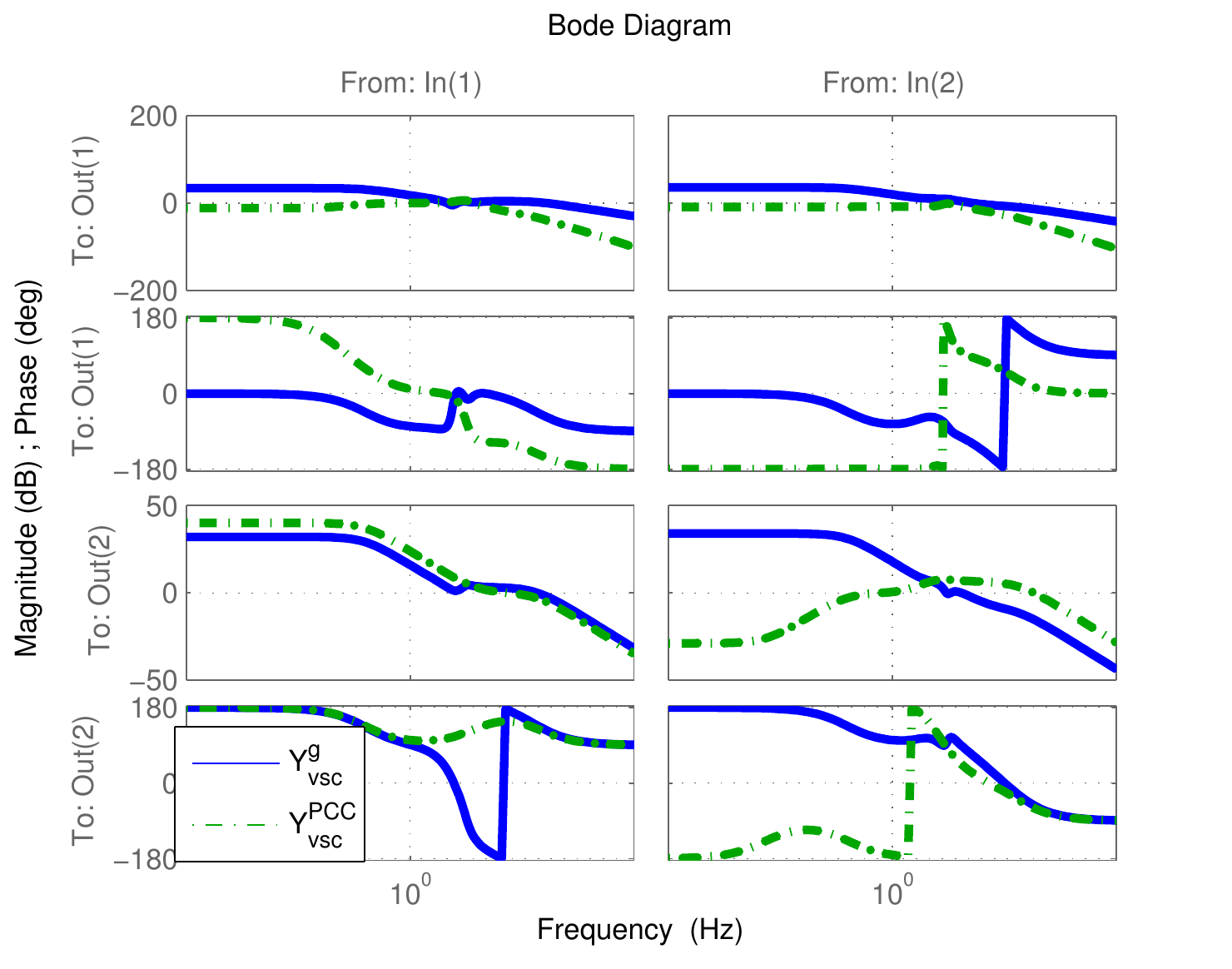}\vspace{-0.1in}
\caption{  \label{fig:vsc_freq1} }
\end{subfigure}
\begin{subfigure}{3.5in}
\centering
\includegraphics[width=3.5in]{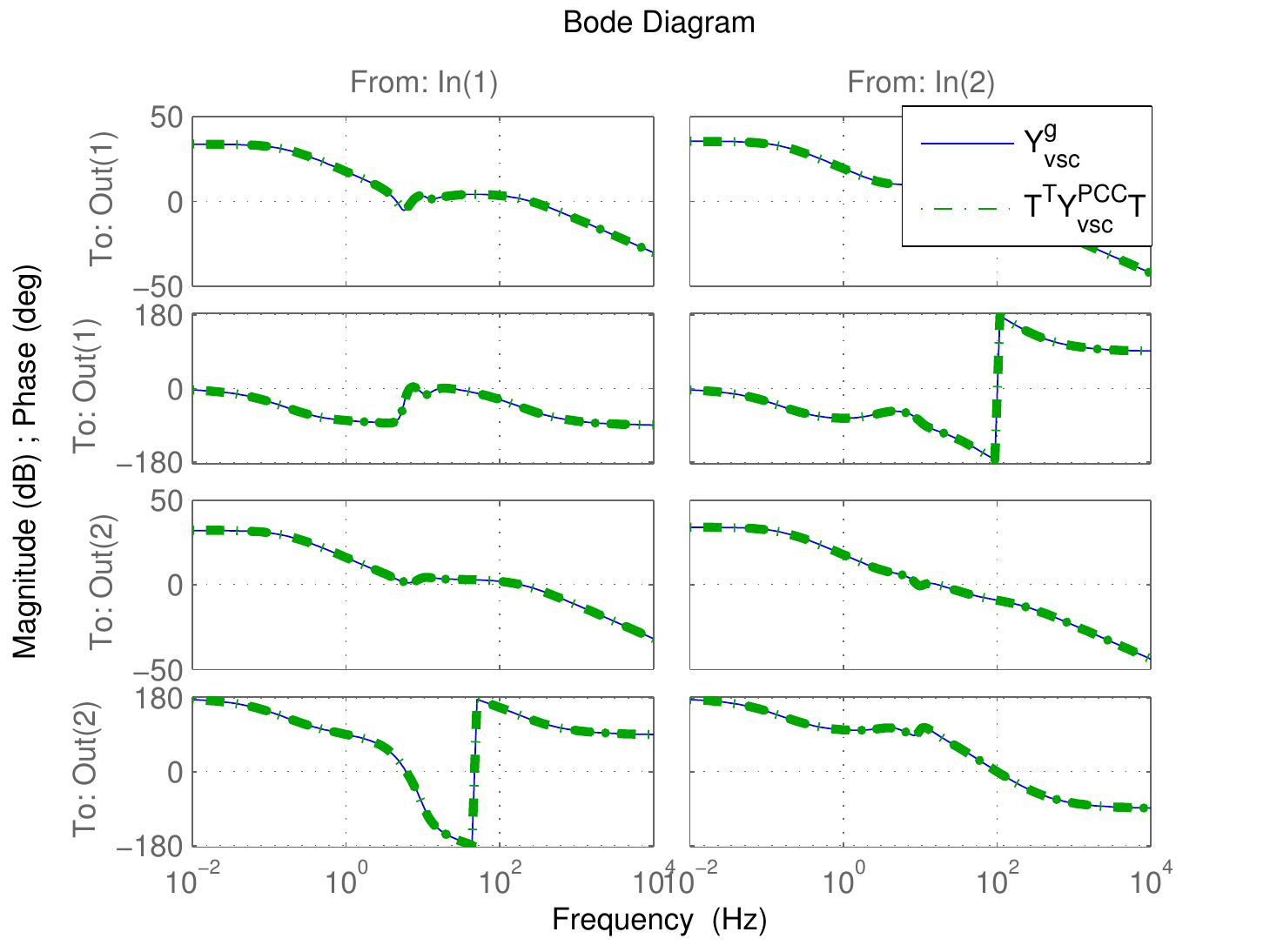}\vspace{-0.1in}
\caption{  \label{fig:vsc_freq2} }
\end{subfigure}
\caption{  (\ref{fig:vsc_freq1}) Bode plots of two admittances $Y^g$ and $Y^{\rm PCC}$  based on two frames. (\ref{fig:vsc_freq2}) Bode plots of  $Y^g$ and $T^TY^{\rm PCC}T$.}
\end{figure}

\subsection{Challenge 2: Reference Frame }
Comparison is also made when different reference frame is adopted during the measurement procedure.
Two reference frames are adopted: grid frame and the PCC frame. For grid frame adoption, the PCC
voltage phasor is $1\phase{50.5^{\circ}}$. For the PCC frame, the PCC voltage phasor is
$1\phase{0}$. It can be seen from Fig. \ref{fig:vsc_freq1} that the admittance frequency domain
responses are very different for different reference frame. On the other hand, based on experiment
results, {\bf the two different admittances lead to the  same eigenvalues}.

This difference in frequency-domain responses and the same closed-loop system eigenvalues  are explained by the following.

The PCC frame leads the grid frame by $\Delta \theta_{\rm PCC}$ (the PCC voltage phase angle
relative to the grid frame). Notate the variables and admittance matrix in the grid frame using
superscript $g$ and those in the PCC frame using superscript $\rm PCC$. The following relationship
can be found.
\begin{align}
I^{\rm PCC} & = Y^{\rm PCC } V^{\rm PCC}, \>\>I^{g}  = Y^{g} V^{g},  \notag\\
I^{\rm PCC} & = T I^{g}, \>\>V^{\rm PCC} = T V^{g},
\end{align}
where \begin{align}T = \begin{bmatrix} \cos(\Delta \theta_{\rm PCC}) & \sin(\Delta\theta_{\rm PCC})\\
-\sin(\Delta\theta_{\rm PCC}) &\cos(\Delta \theta_{\rm PCC})
\end{bmatrix}.\notag\end{align}
$T$ is an orthogonal matrix that defines rotation. Its inverse is its transpose. Hence, it can be
found:
\begin{align} Y^g = T^{-1}Y^{\rm PCC }T,\end{align} where $T^{-1} = T^T$. Fig. \ref{fig:vsc_freq2} confirms the relationship between the two admittances in different frames.

For this particular system, where a single VSC is connected to a voltage source through a transmission line,
 the two different admittances lead to the same eigenvalues.  A question naturally arise is if this
is a general case or just a special case. In general, does reference frame matter? This particular
example is indeed a special case. This is due to the fact that the admittance of the passive
component keeps intact.
\begin{align}
Y_{\rm line} = T^{-1} Y_{\rm line} T.
\end{align}

The total admittances for the two cases are examined:
\begin{align*}
Y^g &= Y^g_{\rm vsc} + Y_{\rm line} \\
&= T^{-1} (Y^{\rm PCC}_{\rm vsc} + Y_{\rm line}) T\\
&= T^{-1} Y^{\rm PCC} T.
\end{align*}

For this specific case, the total admittances in different reference frames are two similar
matrices. They share the same determinant. Hence, the eigenvalues computed are the same.

{\bf Example 7: Two-area four-machine power grid reference frame test} The two-area four-machine
power grid is used as the test case to further investigate the effect of reference frame. In case
1, the system reference frame is adopted for every generator's admittance calibration. This is the
same assumption that is used in Section III's Example 2 eigenvalue computation. In case 2,
component admittance calibration is conducted based on the terminal bus. That is, each generator's
terminal bus's voltage space vector is assumed as the $d$-axis. This leads to $0$ for the $q$-axis
terminal voltage at steady-state.

\begin{figure}[!ht]
\centering
\includegraphics[width=2.5in]{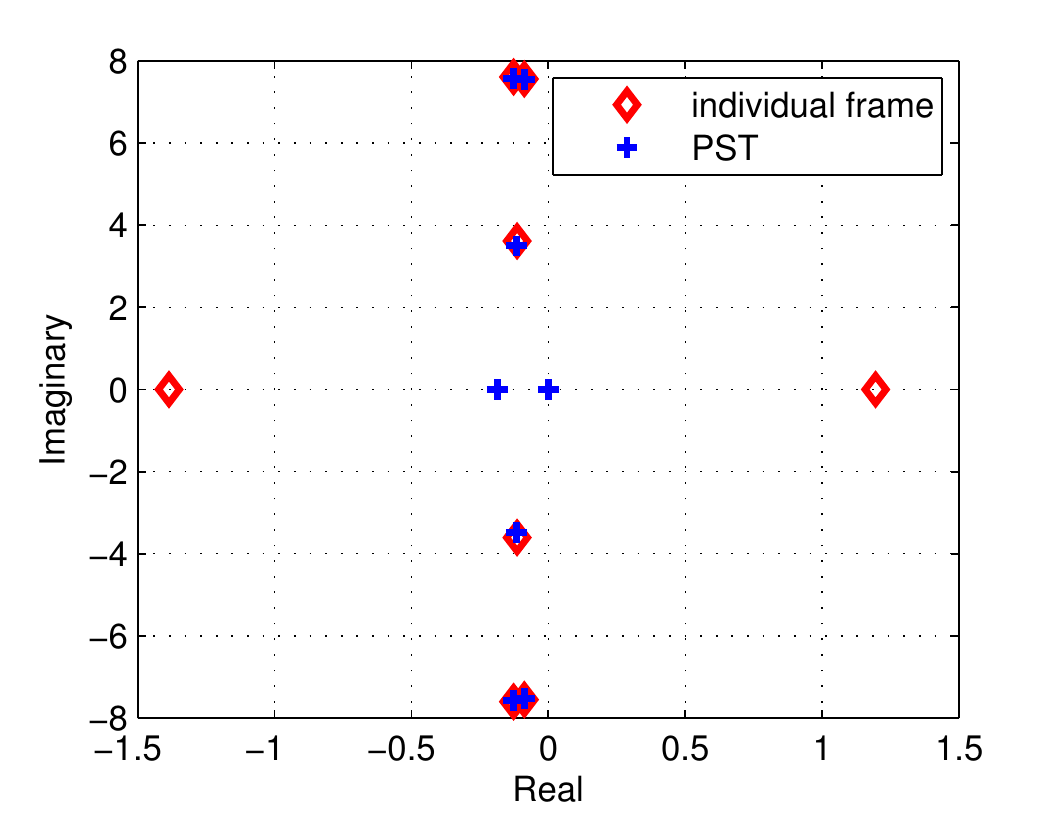}\vspace{-0.1in}
\caption{  Inaccurate eigenvalues of the 2-area 4-machine system due to individual reference frame assumption.}
\label{fig:frame_2a4m}
\end{figure}

The resulting eigenvalues are compared with the PST output and shown in Fig. \ref{fig:frame_2a4m}.
It can be seen that using individual frames leads to inaccurate eigenvalues. Hence, in general, a
system reference frame is desired to lead to accurate network admittance matrix assembling. This
point has also been emphasized in \cite{liu2018oscillatory}.

\subsection{Remarks on Vision 2 }
In Vision 2, two important challenges need to be tackled when forming an network admittance matrix.
First, steady-state analysis of the entire grid with IBR should be conducted. This step is
essential for traditional power grid dynamic stability analysis.  The testbed for admittance
measurement should have the exact same terminal condition as that from the entire grid. This
condition guarantees that a component measured is exactly working at the specific operating
condition.

Second, a system reference frame should be used. If a component is measured at its terminal voltage reference frame, the obtained admittance should be referred back to the system framework before final assembling. 



\section{Vision 3: Aggregation}
Instead of conducting series/shunt branch aggregation as
\cite{liu2017stability, liu2018oscillatory}, efficient and systematic aggregation methods
are desired. In power network steady-state computing, aggregation technology is well established.
One such is Thevenine equivalent. For a network, we may find its Thevenin equivalent impedance at
any bus. The Thevenin equivalent impedance at Bus $k$, notated as $Z_{kk}$ is indeed the $k$th
diagonal component of the inverse of the admittance matrix \cite{bergen1999power} (Chapter 9).

For the frequency-domain network admittance matrix matrix $Y(s)$, aggregation may be conducted at
any frequency for $Y(j\omega)$. Indeed, this aggregation method is exactly the one adopted by Undrill and Kostyniak in \cite{Undrill1976}!

\begin{figure}[!ht]
\centering
\includegraphics[width=2.0in]{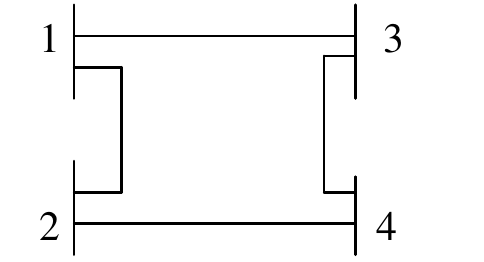}\vspace{-0.1in}
\caption{  Passive network topology of the 2-area 4-machine system after Kron reduction.
The steady-state complex admittance matrix is notated as $Y_{\rm bus}$, a $4\times 4$ matrix.}
\label{ex6}
\end{figure}

{\bf Example 8} In the following, we use the two-area four-machine system as an example to
illustrate how to conduct aggregation. We would like to view the entire grid as two impedances. The
view point is Gen 1's terminal bus. Through Kron reduction and admittance computing, we first
obtain a 4-node network admittance matrix. The topology is shown in Fig. \ref{ex6}.

To find the Thevenine equivalent of the rest system other than Gen 1,  the network admittance
matrix should be built by excluding Gen 1's admittance.
\begin{align}
Y'  = \begin{bmatrix}\mathbf{0} & & & \\ & Y_{g2} & & \\ & & Y_{g3} & \\ & & &Y_{g4}  \end{bmatrix} +
 \begin{bmatrix} [Y_{11}] & [Y_{12}] & [Y_{13}] & [Y_{14}]\\
 [Y_{21}] & [Y_{22}] & [Y_{23}] & [Y_{24}]\\
 [Y_{31}] & [Y_{32}] & [Y_{33}] & [Y_{34}]\\
 [Y_{41}] & [Y_{42}] & [Y_{43}] & [Y_{44}]
\end{bmatrix}\notag
\end{align}
where $[Y_{ij} ]= \begin{bmatrix} \Re(Y_{{\rm bus},ij} ) & -\Im(Y_{{\rm bus},ij} )\\
\Im(Y_{{\rm bus},ij} )& \Re(Y_{{\rm bus},ij} )
\end{bmatrix}$

The first diagonal component  of the inverse of the network admittance matrix $Y'^{-1}$ for the
system without Gen 1 is found, notated as $Z_{11}$. Finally, the total admittance is a $2\times 2$
matrix: $Y = Y_{g1}+ Z^{-1}_{11}$. Numerator of the determinant has 8th order. The roots are found
and they exactly match the eigenvalues of the system computed by PST.

%

\section{Vision 4: Stability Assessment }

The assembled system admittance matrix assumes two formats: transfer function matrix or
frequency-domain measurements. Many methods are suitable only for the first format, including the
eigenvalue approach relying on admittance matrix's determinant's roots.

A few frequency-domain methods, such as Bode plots, reactance crossover,  are suitable for
frequency-domain measurements. Those methods have been popularly used in the literature, e.g.,
\cite{sun2011impedance,cheng2013reactance, cheng2019}. To be able to apply for Bode plots, two
subsystems should be identified. For example, in \cite{cheng2019}, SSR screen is conducted using
reactance cross over method. The system is viewed at the terminal of a series compensated line or a
series capacitor. One subsystem is the capacitor, the other subsystem is the rest system.

Similarly, in \cite{liu2018oscillatory}, an impedance network is presented with every node
considered. The system is then viewed as two subsystems based on the wind farm terminal bus: wind
farm versus the other the rest system. For the rest system, aggregation is conducted.

The two-port system or two-impedance approach has to guarantee that the instability happens only due to the specific two subsystem
interaction. If one subsystem is unstable, the approach is not applicable. An  example of inaccurate analysis is
presented in \cite{cheng2019} to show that if the system is viewed as a single wind farm
versus the rest of the grid, which includes other wind farms and the series compensated line, the
analysis result is not accurate.

For a power grid with multiple nodes, the two-subsystem approach which requires aggregation has two
pitfalls. First, we need to guarantee that the subsystem is stable. This requires prior knowledge
or additional study. Second, with aggregation, information on those aggregated part cannot be
found. Thus, methods suitable to deal with admittance matrices of multiple nodes are desired.

In this regard, RMA that was proposed for harmonic study \cite{xu2005harmonic}, is suitable for the
new need. At every frequency, eigenvalue decomposition may be conducted for the admittance matrix.
Since $V = Y^{-1}I$, singularity of the admittance matrix in the frequency domain will be
manifested as a small eigenvalue of $Y$ and a large eigenvalue of $Y^{-1}$. The eigenvalues of
$Y^{-1}$ are named as modal impedance in \cite{xu2005harmonic}.


If a pair of the system's eigenvalues, notated as $-\sigma \pm j\omega$ have  $\sigma$ very close
to $0$, the determinant of $Y(s)$, evaluated at $j\omega$ will reach $0$. This indicates
singularity of $Y(j\omega)$ and in turn a large modal impedance at this frequency
$\omega$. Singularity of $Y$ can also be manifested as the minimum of singular values of $Y$ becomes very small.
MATLAB's command \texttt{sigma} gives the singular value plots over frequency for a transfer function matrix.

In the following, we present three examples to demonstrate stability assessment.

 The VSC in weak grid
example ({\bf Example 6}) is again used to illustrate RMA and sigma plots. In the RMA analysis, we
deal with an admittance matrix $Y = Y_{\rm vsc}  + Y_{\rm line}$. The eigenvalues of the matrix in
the frequency domain are computed and the inverse of the eigenvalues (modal impedances) are plotted
in Fig. \ref{Fig:VSC_RMA}. Since the matrix is  $2\times2$, the system has two modal impedances.
One of the modal impedances show a large value in the range of 5 Hz -6 Hz. When the grid strength
is weaker or the line reactance is larger, the maximum of the modal impedance becomes even larger.
The singular value plot of $Y$ is also shown in Fig. \ref{Fig:VSC_RMA}. It can be seen that longer
line leads to smaller minimum singular value, indicating the system is approaching singularity.

%

\begin{figure}[!ht]
\centering
\begin{subfigure}{1.72in}
\includegraphics[width=1.73in]{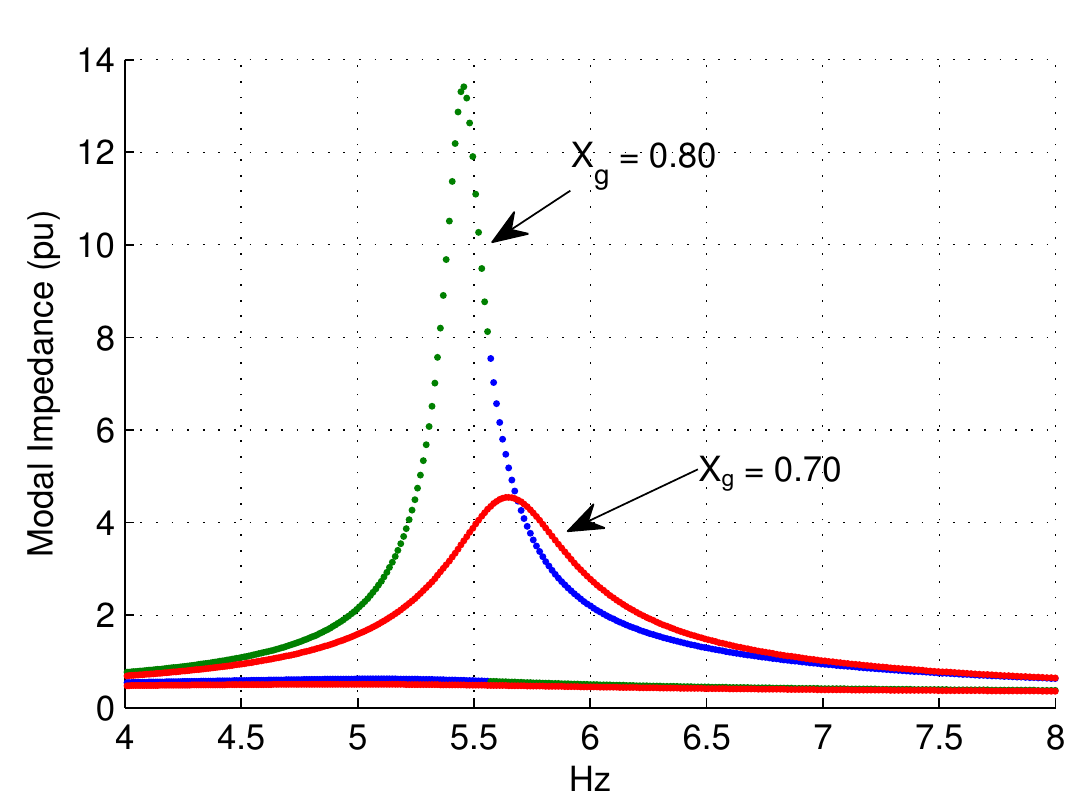}
\end{subfigure}
\begin{subfigure}{1.72in}
\includegraphics[width=1.73in]{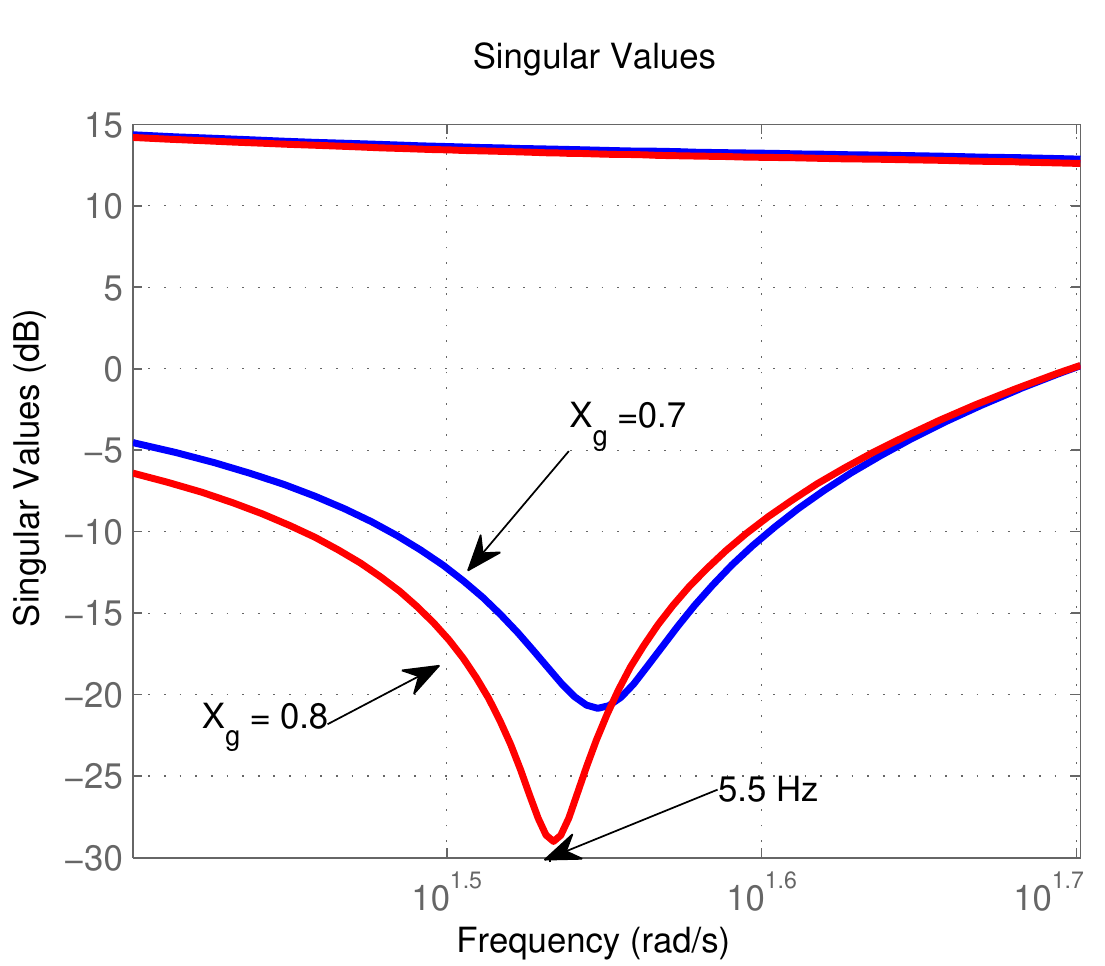}
\end{subfigure}
\caption{ (a) Modal impedances for VSC in weak grids. A larger modal impedance is observed when line is longer. The resonance happens at 5.5 Hz.
(b) Singular value plot for the admittance matrix. Longer line leads to smaller minimum singular value.  }
\label{Fig:VSC_RMA}
\end{figure}

{\bf Example 9: Subsynchronous resonance due to induction generator effect} This example is taken
from \cite{fan2012nyquist}. A type-3 wind turbine is radially connected to a series compensated
network. To arrive at frequency-domain impedance model, the slip should be represented by
$\text{slip} = 1-\frac{j\omega_m}{s}$, where $\omega_m$ is the motor speed in rad/s.

\begin{figure}[!ht]
\centering
\includegraphics[width=3.2in]{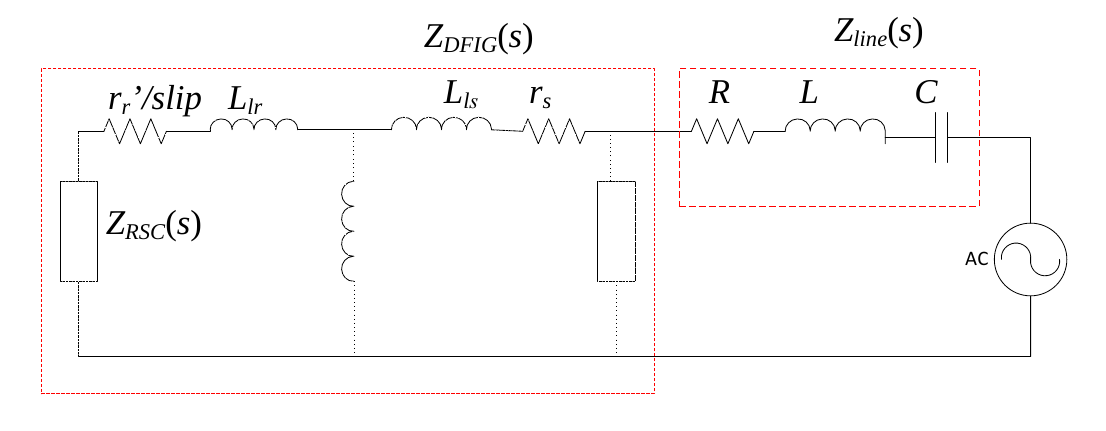}\vspace{-0.1in}
\caption{ Equivalent circuit of a type-3 wind connected to an RLC circuit. $r_s, X_{ls}, r_r, X_{lr}$: $0.00488$,
$0.09231$, $0.00549$, $0.09955$. $\omega_m=0.75$ pu. $R, X_L$: $0.03, 0.64$. }
\label{circuit_type3ssr}
\end{figure}
\begin{figure}[!ht]
\centering
\includegraphics[width=2.5in]{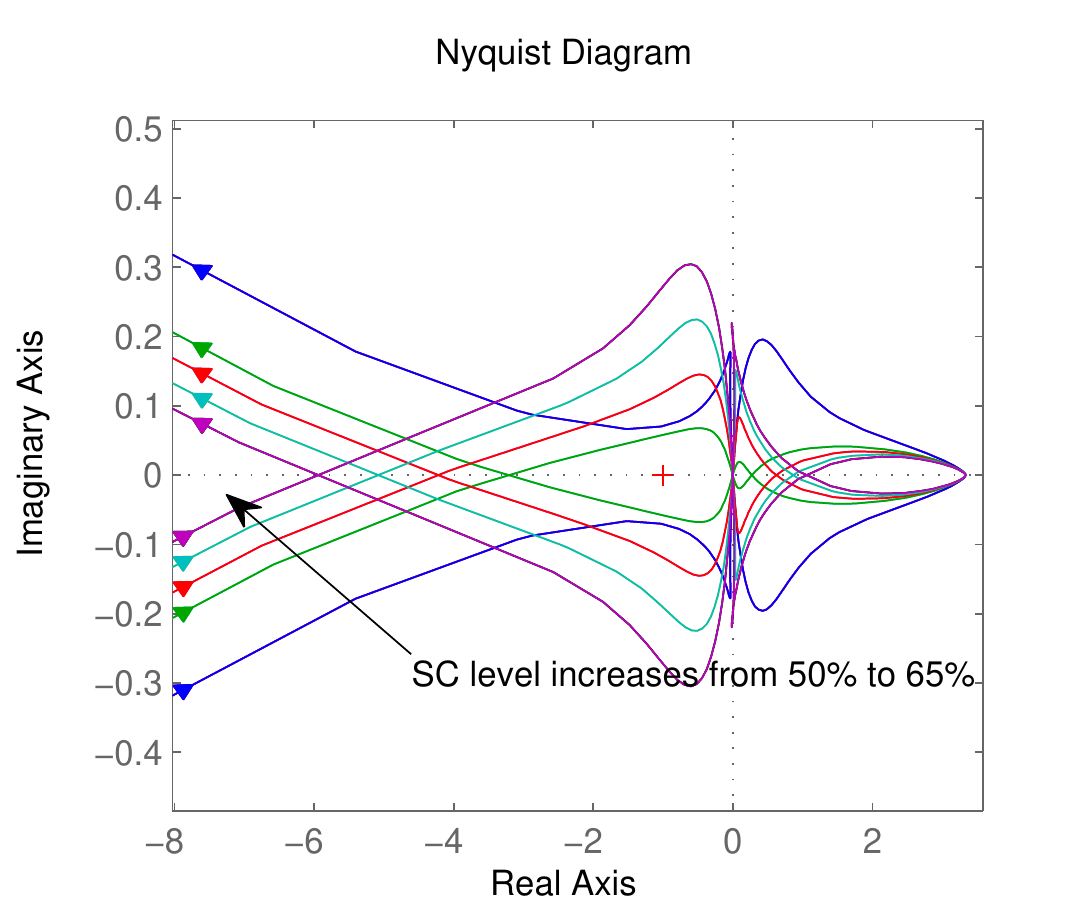}
\caption{  Nyquist plots of $Y_{\rm DFIG}Z_{\rm Line}$. SC at 50\%: (-1,0) is not encircled clockwise and the system is stable. SC at 57.5\%, 60\%, 62.5\%, 65\%: (-1,0) encircled unstable. }
\label{Fig:DFIGSSR_Nyquist}
\end{figure}

Impedance or admittance-based frequency domain analysis can provide better insights for this
dynamic phenomenon. The equivalent circuit is shown in Fig. \ref{circuit_type3ssr}.
 For a doubly-fed induction generator (DFIG), the shunt branches of grid side converter (GSC) and the
magnetizing branch are assumed to have large impedance and thus treated as open. The rotor side
converter (RSC)'s current control gains are assumed very small so that the resulting impedance can
also be ignored. Nyquist plots of $Y_{\rm DFIG}Z_{\rm line}$ have been presented in
\cite{fan2012nyquist}. They are given in Fig. \ref{Fig:DFIGSSR_Nyquist} to demonstrate the effect
of series compensation (SC) level on resonance. It can be seen that for the series compensation
levels invesigated, 50\% does not lead to instability while the rest lead to instability.

We also compute the roots of the total admittance in \eqref{eq:ssr} in the static frame.
\begin{equation}
\begin{aligned}
Y &= Y_{\rm DFIG} + Y_{\rm Line} \\
 & =\frac{1}{r_r/\text{slip} + r_s + (L_{ls}+L_{lr})s} +\frac{1}{R+Ls+1/Cs}
\end{aligned}
\label{eq:ssr}
\end{equation}

\begin{figure}[!ht]
\centering
\includegraphics[width=2.5in]{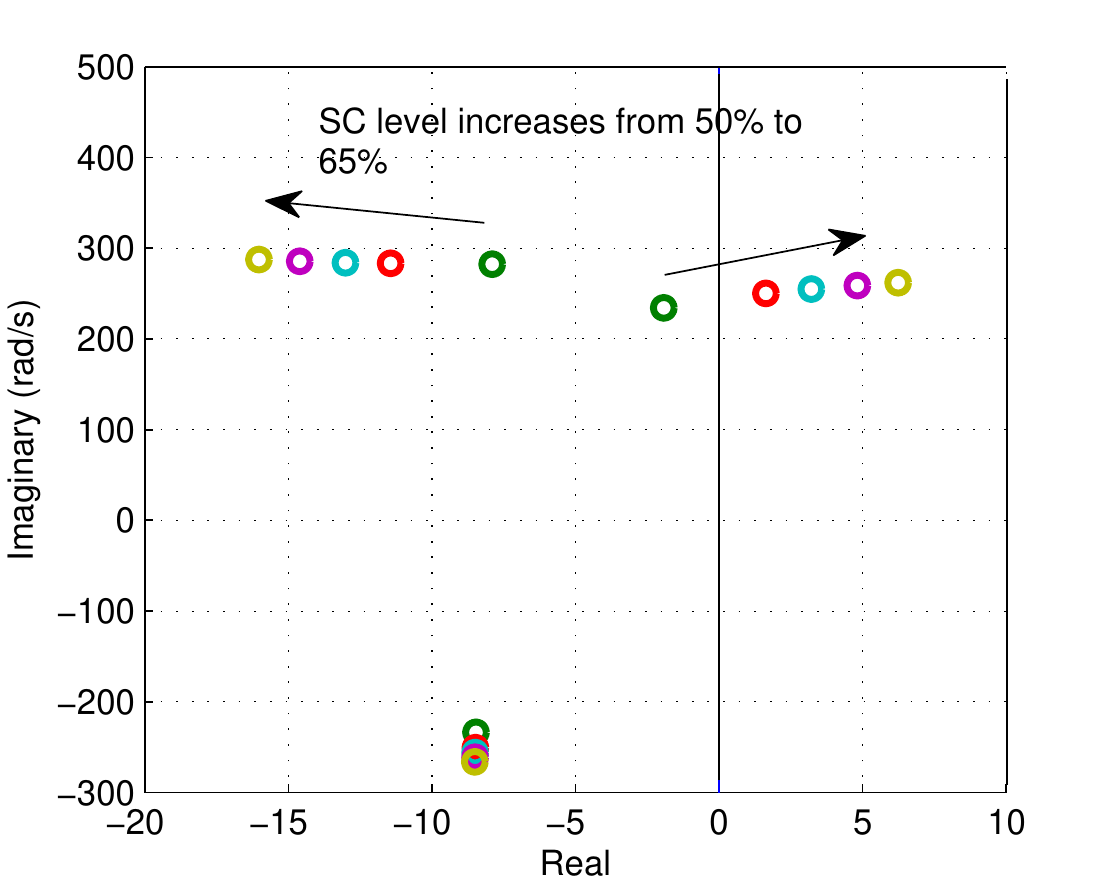}
\caption{  Eigenvalues of the system in the static frame show a dominant mode with oscillation frequency in the range of 37 Hz to 45 Hz when SC changes: 50\%, 57\%, 60\%, 62\%, 65\%.}
\label{Fig:DFIGSSR_eig}
\end{figure}

Fig. \ref{Fig:DFIGSSR_eig} presents the eigenvalues of the system. A dominant mode in the range of 37 Hz to 45 Hz appears when the series compensation level increases from 50\% to 65\%. The eigenvalue plot confirms that the system is stable at 50\% compensation level and unstable at the rest compensate levels.

Next, RMA is conducted. Fig. \ref{Fig:VSC_RMA} presents the modal impedance for the scalar admittance at different compensation level.
It can be seen that at 56\% compensation level, the maximum of the modal impedance achieves a great value.
This is due to the fact that at 56\% compensation level, the dominant mode has the smallest real part.
Hence, at the corresponding mode frequency, the magnitude of the admittance is very small or the modal impedance is very large.
 RMA is most effective to indicate that a system has an oscillation mode at a certain frequency. On the other hand, RMA does not tell if the system is stable or not.
  When the compensation level changes from 50\% to 56\%, if the system is previously stable, the sudden increase of the modal impedance magnitude indicates the system may start to lose stability.

\begin{figure}[!ht]
\centering
\includegraphics[width=3.0in]{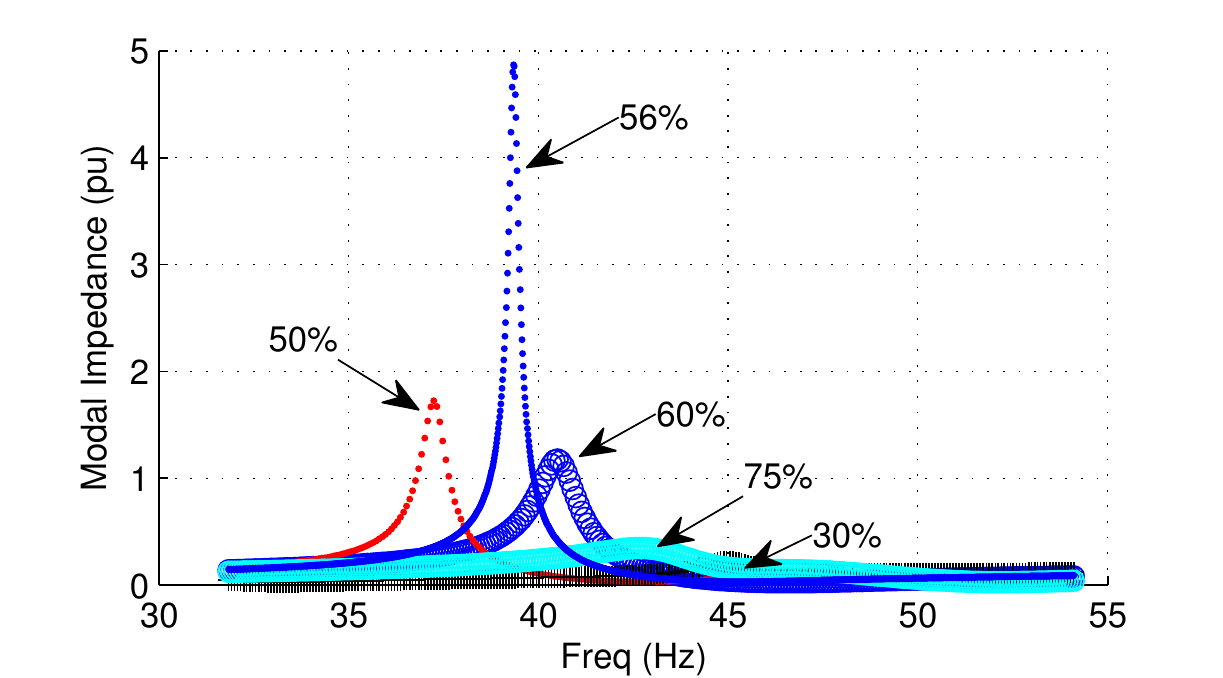}
\caption{  Modal impedance for Example 9. Modal impedance is maximum when the system's eigenvalue is close to the imaginary axis.}
\label{Fig:VSC_RMA}
\end{figure}

%
%

\begin{figure}[!ht]
\centering
\begin{subfigure}{1.72in}
\includegraphics[width=1.72in]{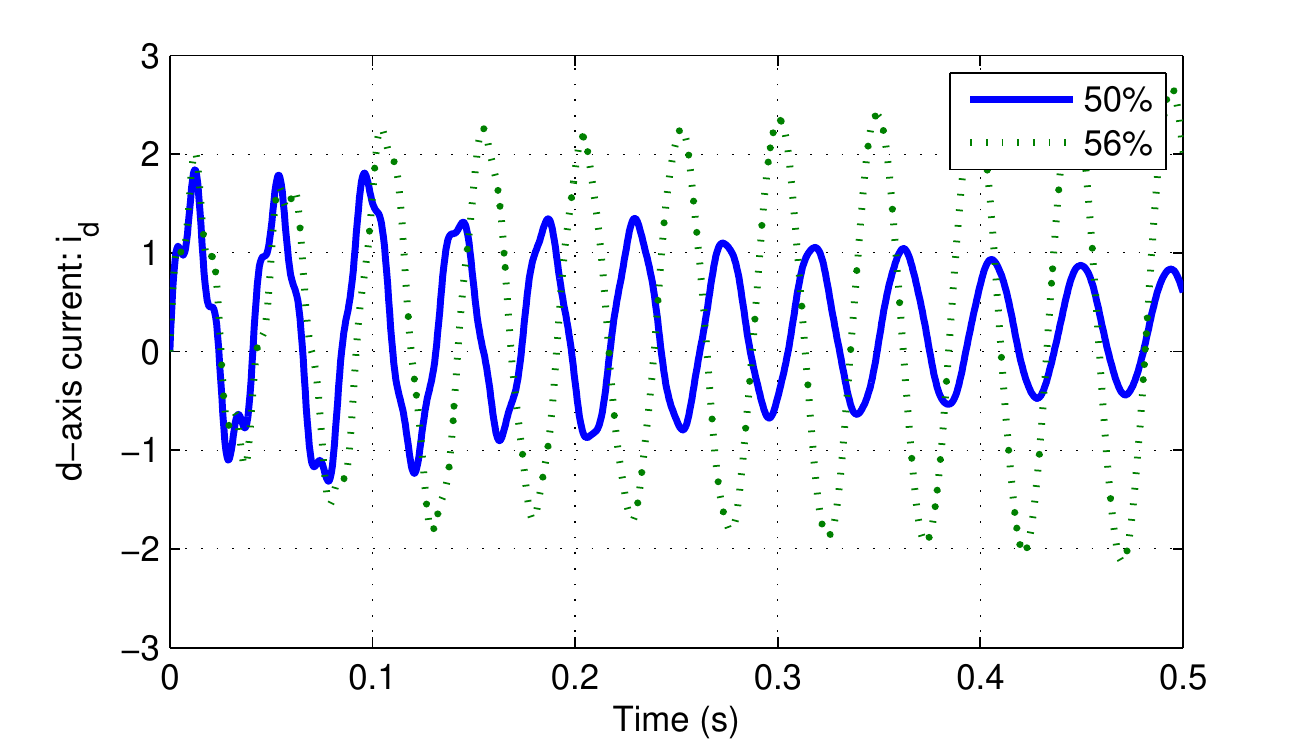}
\caption{  \label{fig:sim_DFIGSSRa}}
\end{subfigure}
\begin{subfigure}{1.72in}
\includegraphics[width=1.72in]{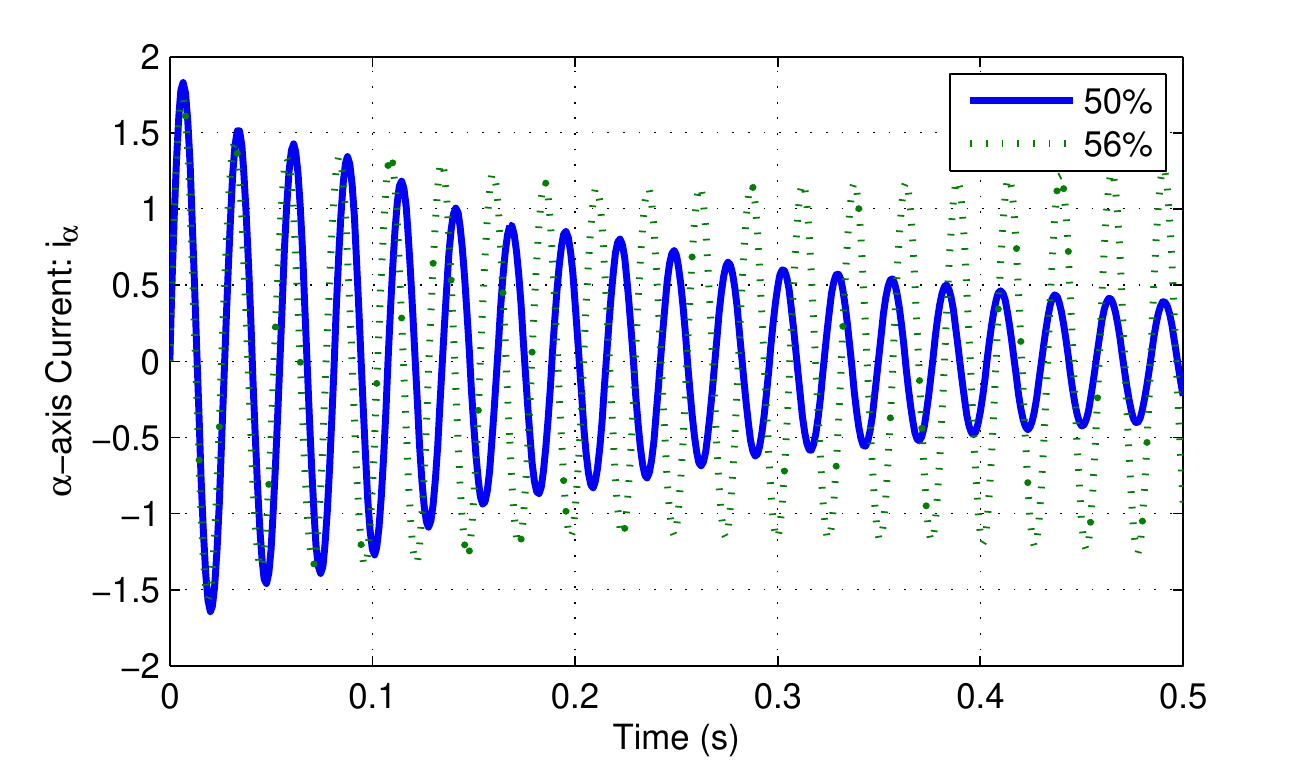}
\caption{  \label{fig:sim_DFIGSSRb}}
\end{subfigure}
\caption{  (\ref{fig:sim_DFIGSSRa}) Step response of $i_d$ subject to a step change in $v_d$ of the grid voltage.
(\ref{fig:sim_DFIGSSRb}) Step response of $i_\alpha$ subject to a step change in $v_\alpha$ of the grid voltage.
SC at 50\%: stable and oscillation frequency is 23 Hz in $dq$, 37 Hz in static frame. SC at 56\%: unstable. }
\label{Fig:DFIGSSR_sim}
\end{figure}

Finally, time-domain step response of current due to step change in grid voltage is examined. The
transfer function, $\frac{1}{Z_{\rm DFIG} + Z_{\rm line}}$, in the static frame has complex terms.
Thus, the transfer function is  converted to $\alpha\beta$ frame and $dq$-frame. For any transfer
function notated as $F(s)$ in the static frame, its transfer function matrix in $\alpha\beta$ frame
and $dq$-frame are shown as follows.
\begin{align}
F_{\alpha \beta}(s) &=\begin{bmatrix} \Re(F(s)) & -\Im(F(s)) \\ \Im(F(s)) & \Re(F(s)) \end{bmatrix}\\
F_{dq}(s) &=\begin{bmatrix} \Re(F(s+j\omega_0)) & -\Im(F(s+j\omega_0)) \\ \Im(F(s+j\omega_0)) & \Re(F(s+j\omega_0)) \end{bmatrix}
\end{align}
Note that $s$ is replaced by $s+j\omega_0$ to find transfer functions in the $dq$-frame. 
 A step change is applied in the $d$-axis voltage or $\alpha$-axis of the voltage, the current responses are shown in Fig.
 \ref{Fig:DFIGSSR_sim}. It can be seen oscillations manifest as 37 Hz in the $\alpha\beta$ static
 frame and $23$ Hz in the $dq$-frame.

\vspace{0.1in} {\bf Example 10: Torsional interaction in a synchronous generator radially connected
to an RLC circuit}

In this last example, we demonstrate how to build admittance model for a synchronous generator with
torsional dynamics included. The study system is a synchronous generator connected to an infinite
bus through an RLC circuit. 
The
generator electric model is assumed to be a constant voltage behind a transient reactance.
Parameters of the torsional dynamics are taken from \cite{kundur1994power}.

The procedure of admittance block building was mentioned in the 1976 paper \cite{Undrill1976} with
Nyquist plots presented for the open-loop gain ($Y_{\rm gen}Z_{\rm line}$). In the current paper,
frequency response Bode plots of the obtained generator admittance with torsional dynamics included
will be presented. This generator admittance and the RLC circuit admittance are assembled into a
total admittance $Y$. Eigenvalues of the system computed from $Y$ and the RMA analysis results will
be presented. As a comparison, a conventional linear model is constructed. The open-loop analysis
using the Root-Locus method  demonstrates a $24$ Hz mode losing stability when the compensation
level is 35\%. 

\begin{figure}[!ht]
\centering
\includegraphics[width=2.8in]{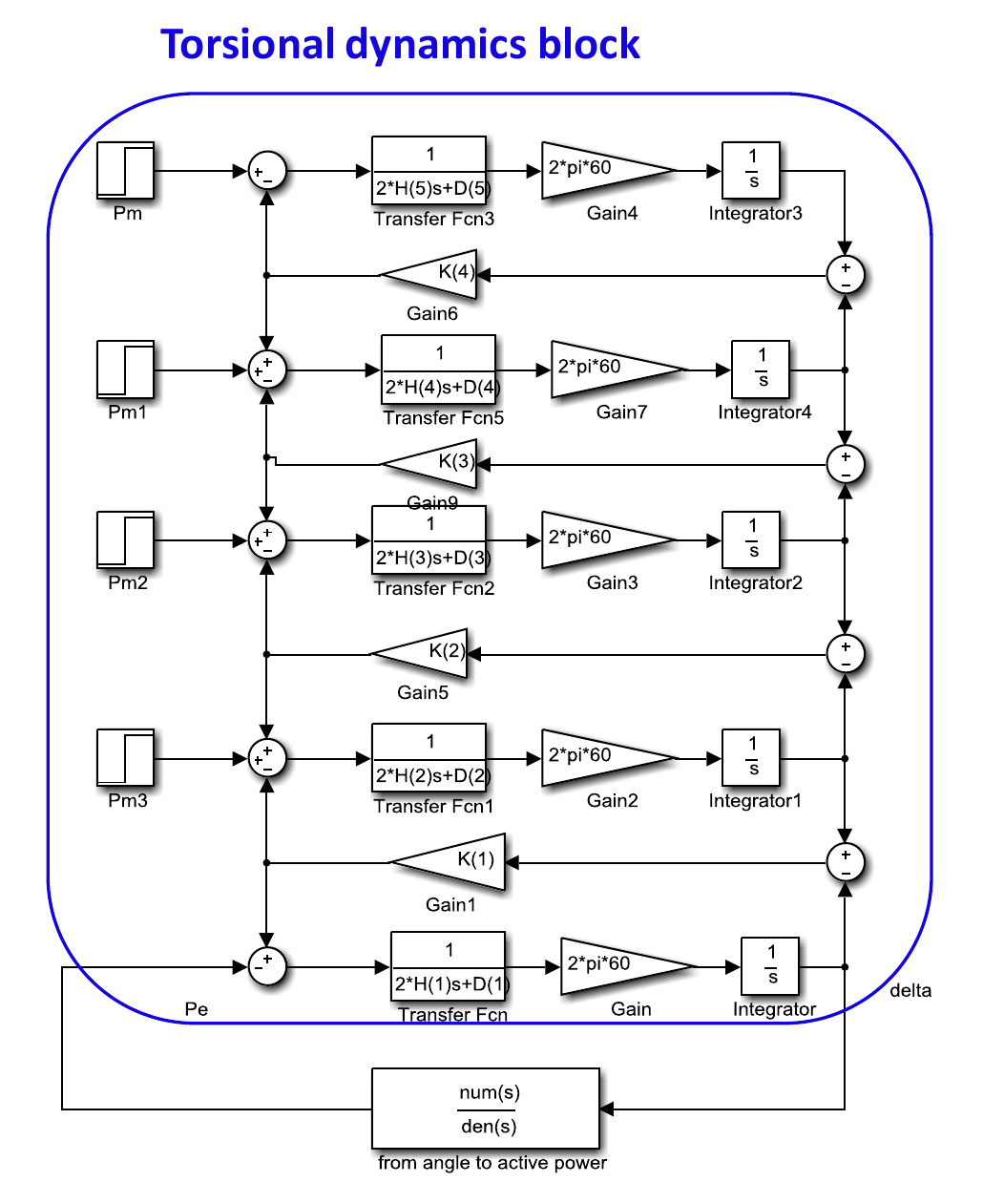}
\caption{Conventional model building blocks in MATLAB/Simulink. System base: 110 MW. RLC circuit parameters: $X_L = 1.0$, $R_L=0.1$, $X_c = 0.35$.
Generator base 600 MW. $H=[0.855 1.192 1.155 0.232 0.124]$, $D =[ 0.3104 0.3104 0.3104 0.05 0.05]$, $K=[62.3 75.6 48.4 21.8]$. Generator operating condition: $P,Q,V$: $0.5, 0.1, 1$.  }
\label{fig:ti1}
\end{figure}

\begin{figure}[!ht]
\centering
\includegraphics[width=2.8in]{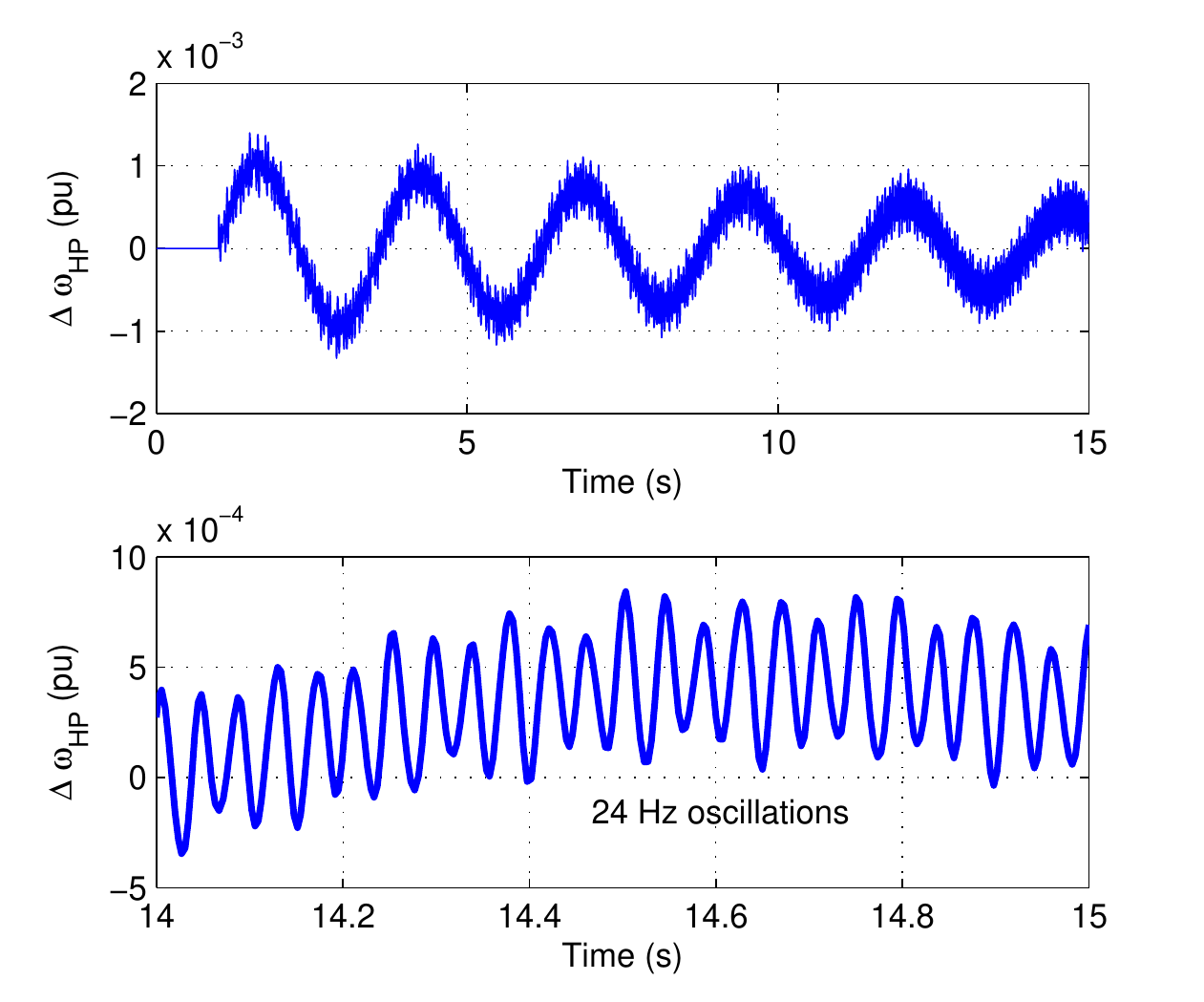}
\caption{  Dynamic performance of the high pressure turbine mass' speed. }
\label{fig:ti2}
\end{figure}

 The conventional linear
model building blocks are presented in Fig. \ref{fig:ti1} with the system parameters listed in the
caption. The mechanical system and the electric system are coupled through the rotor angle $\delta$
and the electric power $P_e$. Linear relationship between $\delta$ and $P_e$ for an RLC circuit has
been derived and presented in the first author's textbook Chapter 8.3 \cite{fan2017control}. At 1
second, a step change is applied at the mechanical power. The resulting dynamic performance of the
speed of the high pressure turbine mass is shown in Fig. \ref{fig:ti2}. It can be seen that there
is a 24 Hz oscillation mode.

\begin{figure}[!ht]
\centering
\includegraphics[width=2.8in]{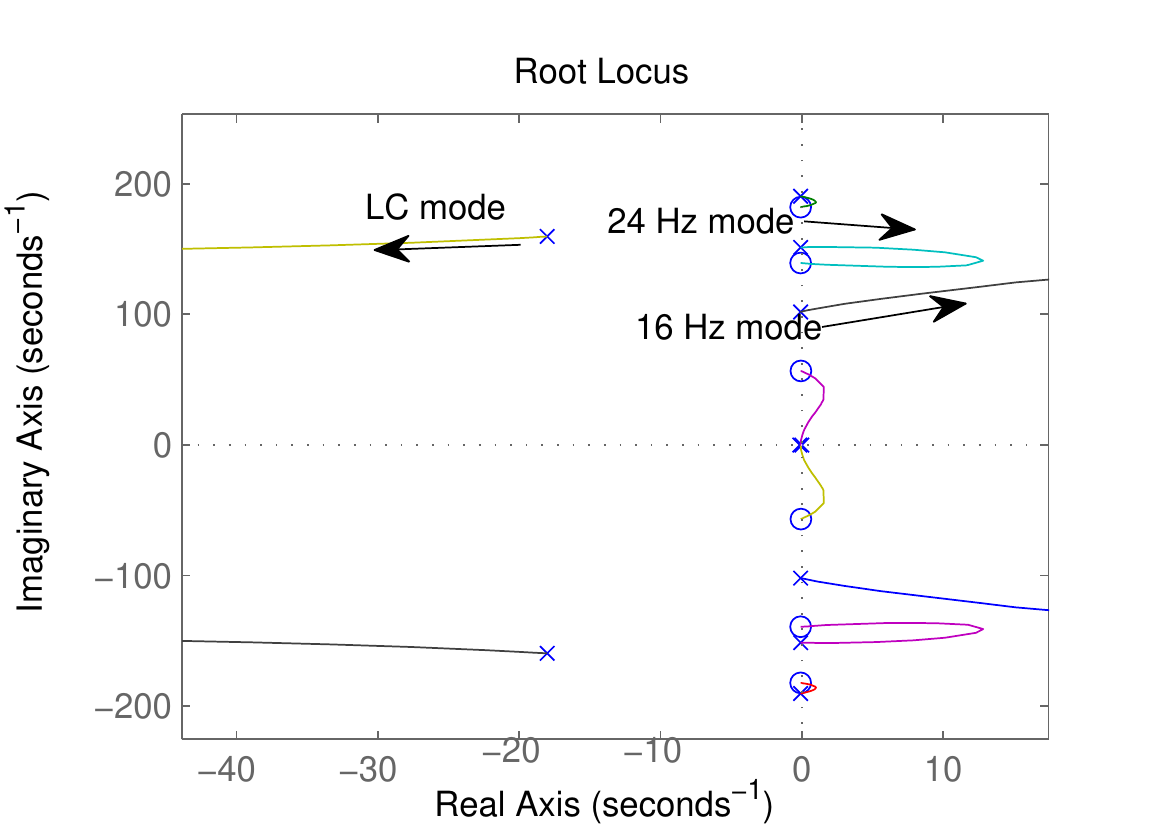}
\caption{  Root loci show two torsional modes (16 Hz, 24 Hz) moving towards RHP. At unit gain, the 24 Hz mode is located in the RHP.  }
\label{fig:ti_rootlocus}
\end{figure}

At the point of $P_e$ that fed to the torsional block, the system is decoupled into an open-loop
system. The root loci are obtained using the Root-Locus method, shown in Fig.
\ref{fig:ti_rootlocus}. It can be seen that closing the loop causes the LC mode at 25.3 Hz moving
left while the two torsional modes at 16 Hz and 24 Hz moving towards the RHP. At unity gain, the 24
Hz is located at the RHP, which indicates instability of the closed-loop system.

\begin{figure}[!ht]
\centering
\includegraphics[width=3.2in]{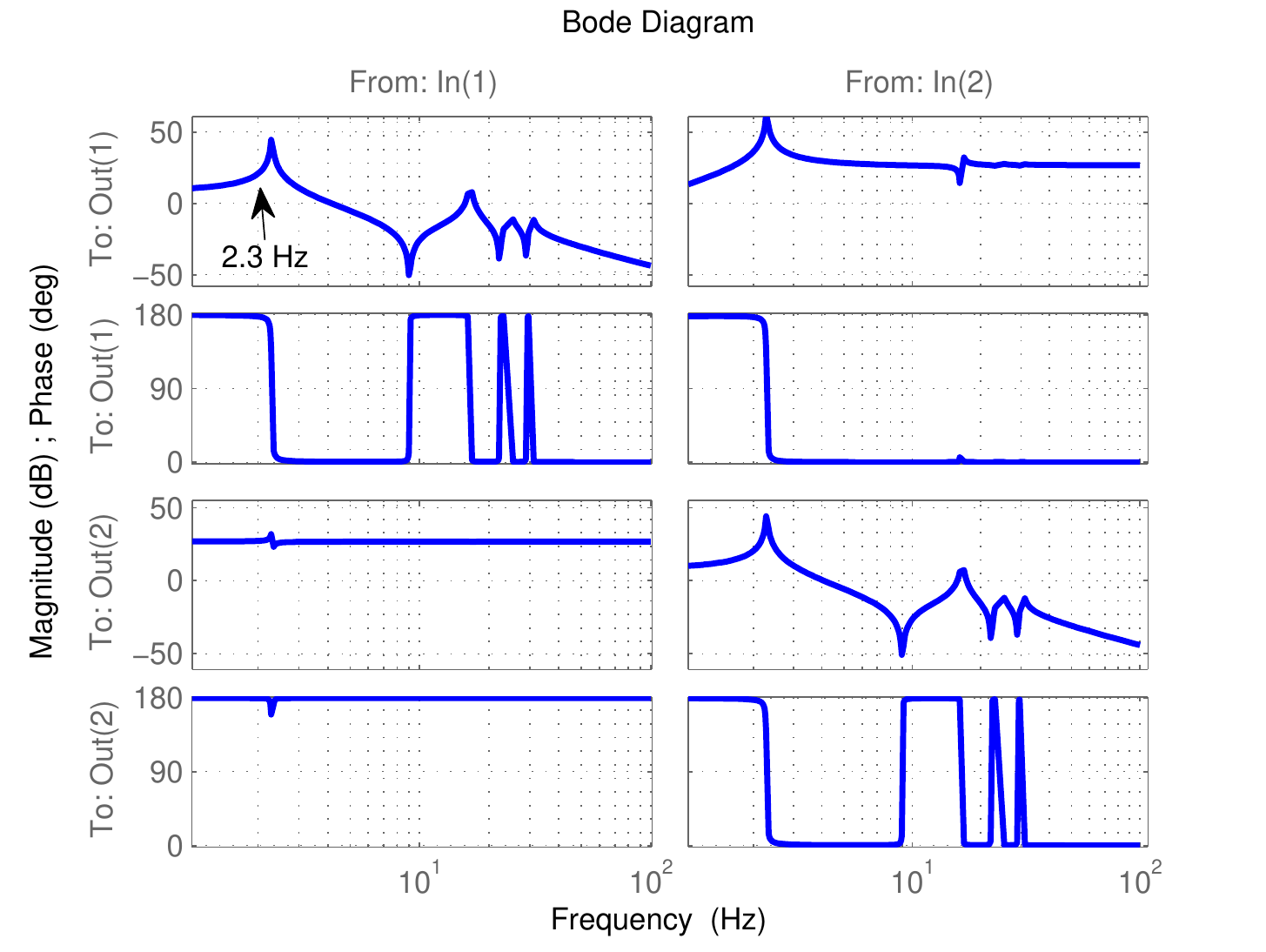}
\caption{  Bode plots of the generator admittance. Peaks in magnitudes correspond to 2.3 Hz, 16 Hz, 24 Hz, and 31 Hz. }
\label{fig:bode_ti}
\end{figure}

The generator's admittance is obtained by considering a generator connected at the terminal bus to
a voltage source. 
Bode plots of the admittance are shown in Fig. \ref{fig:bode_ti}.


\begin{figure}[!ht]
\centering
\includegraphics[width=2.8in]{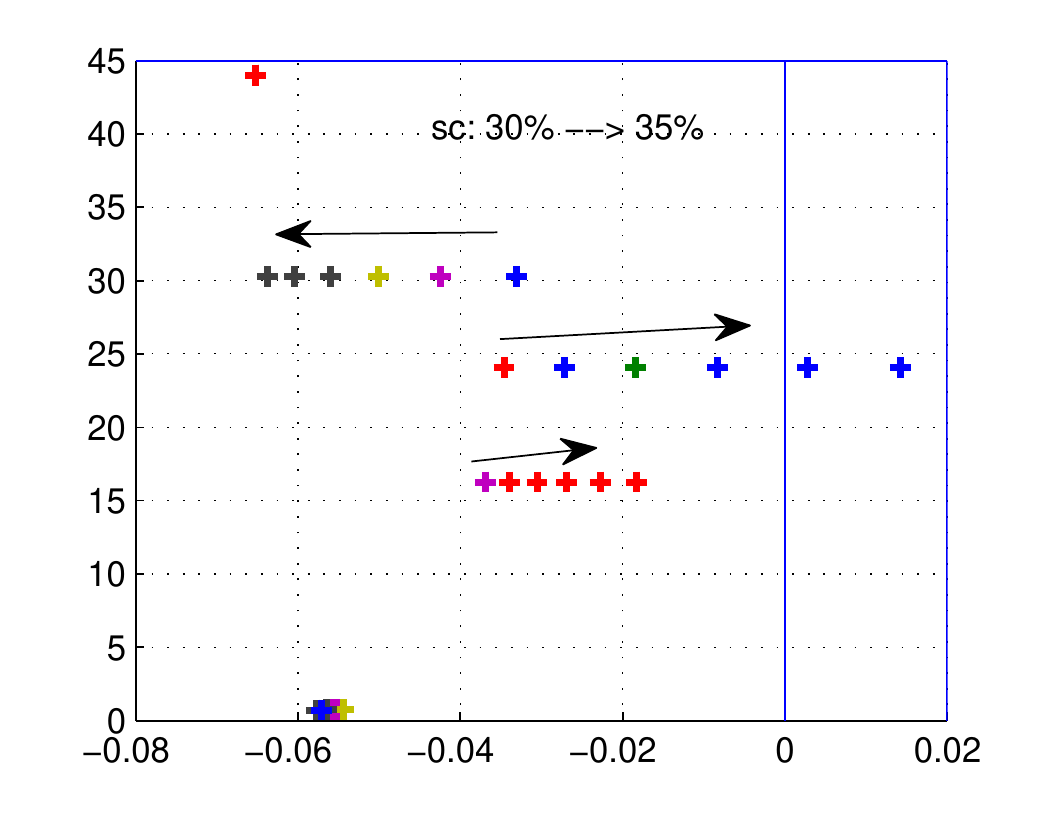}
\caption{  System eigenvalues computed using total admittance $Y$.}
\label{fig:ti_roots}
\end{figure}

The eigenvalues of the system subject to a increasing series compensation are computed using the
assembled $Y$, shown in Fig. \ref{fig:ti_roots}. It can be seen that the 16 Hz mode 24 Hz mode are
moving towards right when series compensation level increases.

\begin{figure}[!ht]
\centering
\includegraphics[width=3.0in]{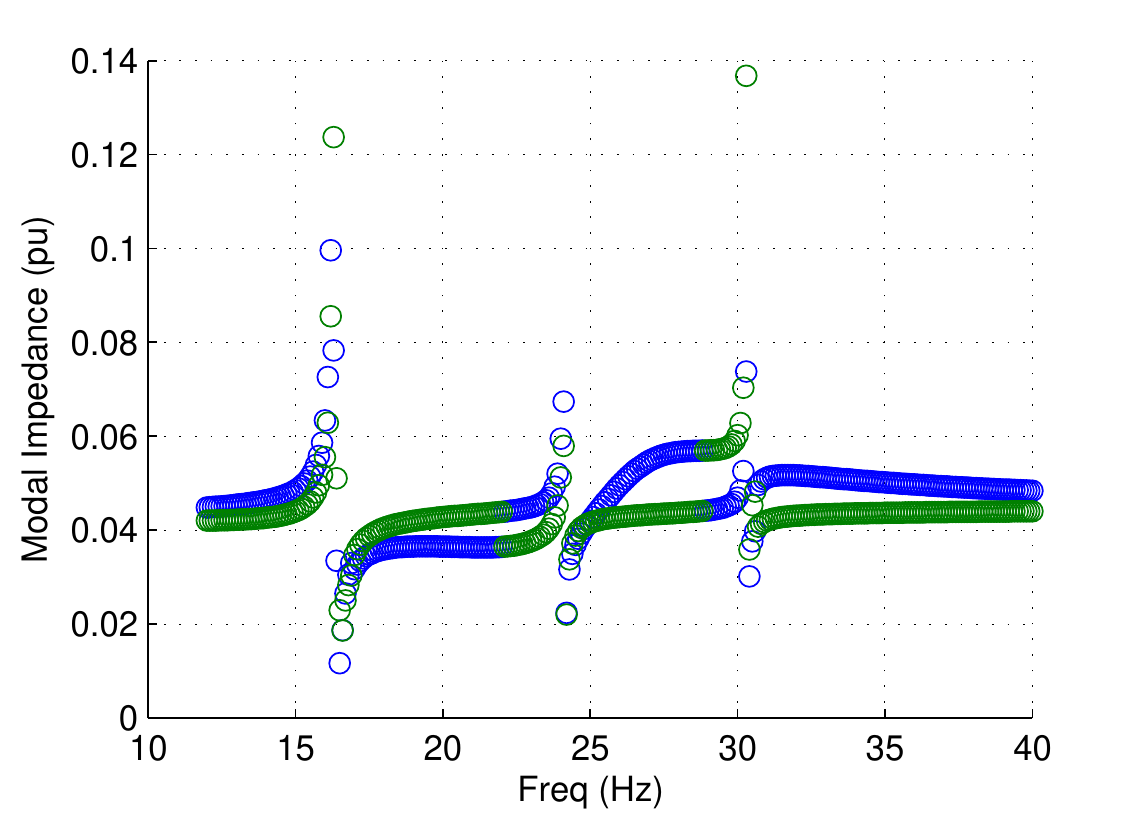}
\caption{ Modal impedances show resonance mode at 16 Hz, 24 Hz, and 31 Hz.  }
\label{fig:TI_RMA}
\end{figure}

Finally, the modal impedances in the frequency domain are computed and shown in Fig.
\ref{fig:TI_RMA}. At the frequencies corresponding to three torsional modes, increase in modal
impedance magnitudes is observed.

\textbf{Remarks:} In Vision 4, stability assessment tools are reviewed. Tools suitable for the
admittance measurement-based network model are desired. In this paper, we use three examples to
demonstrate the use of roots of $\det(Y)$ and RMA for stability or robustness assessment. Further
investigation implementing those methods on large size systems is desired. Further, tool such as
participation factor that have been used in traditional analysis to reveal component involvement in
oscillation modes need to be re-designed for the new framework.

\section{Conclusion}
This visionary paper lays out visions for the modular small-signal analysis framework suitable for
power grids with IBR penetration.  The four visions are related to component admittance
measurement, network admittance matrix assembling, aggregation, and stability assessment. First, a
stability criterion suitable for a general network admittance matrix is  presented. Two approaches
to obtain an admittance are then presented: the white-box approach and the black-box
approach. For both approaches, efficient methods are examined. 
Further, this paper uses numerical examples to demonstrate challenges in network admittance matrix
assembling and how to conduct efficient aggregation. Finally, stability assessment is carried out
on examples using various methods: Nyquist criterion, roots of the determinant of the admittance
matrix, and RMA method.

The visions as well as the numerical examples are expected to  provide guidelines and references for a modular modeling and small-signal analysis framework.

\section*{Acknowledgement}
The authors wish to acknowledge the editor-in-chief Prof. Wilsun Xu and two anonymous reviewers for
providing insightful and detailed suggestions on our visionary paper proposal.

\bibliographystyle{IEEEtran}
\bibliography{IEEEabrv,bib}
\vspace{-0.4in}
\begin{IEEEbiographynophoto}{Lingling Fan}(SM'08) received the B.S. and M.S. degrees in electrical engineering from Southeast
University, Nanjing, China, in 1994 and 1997, respectively, and the Ph.D. degree in electrical
engineering from West Virginia University, Morgantown, in 2001.
Currently, she is an Associate Professor with the University of South Florida, Tampa, where she has
been since 2009. She was a Senior Engineer in the Transmission Asset Management Department, Midwest
ISO, St. Paul, MN, form 2001 to 2007, and an Assistant Professor with North Dakota State
University, Fargo, from 2007 to 2009. Her research interests include power systems and power
electronics. Dr. Fan serves an editor for IEEE Trans. Sustainable Energy.
\end{IEEEbiographynophoto}
\vspace{-0.3in}
\begin{IEEEbiographynophoto}
{Zhixin Miao}(SM'09)
received the B.S.E.E. degree from the Huazhong University of Science and Technology,Wuhan, China,
in 1992, the M.S.E.E. degree from the Graduate School, Nanjing Automation Research Institute
(Nanjing, China) in 1997, and the Ph.D. degree in electrical engineering from West Virginia
University, Morgantown, in 2002.

Currently, he is with the University of South Florida (USF), Tampa. Prior to joining USF in 2009,
he was with the Transmission Asset Management Department with Midwest ISO, St. Paul, MN, from 2002
to 2009. His research interests include power system stability, microgrid, and renewable energy.
\end{IEEEbiographynophoto}

\end{document}